\documentclass{ws-ijmpa}
\usepackage{amssymb,graphicx,amsmath,graphics,xfrac,url,xcolor,booktabs}

\newcommand\D{\mathrm{d}}

\begin{document}

\title{The DKP Equation in Presence of a Woods--Saxon Potential: Transmission Resonances and Bound States}
\author{Alexis Garz\'on}
\address{Yachay Tech University, School of Physical Sciences and Nanotechnology, Hda. San Jos\'e S/N y Proyecto Yachay, 100119, Urcuqu\'i, Ecuador.\\alexis.garzon@yachaytech.edu.ec
}

\author{Ricardo Fern\'andez}
\address{Yachay Tech University, School of Physical Sciences and Nanotechnology, Hda. San Jos\'e S/N y Proyecto Yachay, 100119, Urcuqu\'i, Ecuador.\\ricardo.fernandez@yachaytech.edu.ec
}

\author{Vicente A. Ar\'evalo}
\address{Yachay Tech University, School of Physical Sciences and Nanotechnology, Hda. San Jos\'e S/N y Proyecto Yachay, 100119, Urcuqu\'i, Ecuador.\\
vicente.arevalo@yachaytech.edu.ec}

\author{David Laroze}
\address{Instituto de Alta Investigaci\'on, Universidad de Tarapac\'a, Casilla 7 D,
1000000--Arica, Chile.\\
dlarozen@academicos.uta.cl
}

\author{Laura M. P\'erez}
\address{Departamento de Ingenier\'ia Industrial y Sistemas, Universidad de
Tarapac\'a, Casilla 7 D, 1000000--Arica, Chile.\\
lperez@academicos.uta.cl
}

\author{Benjam\'in de Zayas}
\address{Yachay Tech University, School of Physical Sciences and Nanotechnology, Hda. San Jos\'e S/N y Proyecto Yachay, 100119, Urcuqu\'i, Ecuador.\\
bdezayas@yachaytech.edu.ec}

\author{Clara Rojas}
\address{Yachay Tech University, School of Physical Sciences and Nanotechnology, Hda. San Jos\'e S/N y Proyecto Yachay, 100119, Urcuqu\'i, Ecuador.\\
crojas@yachaytech.edu.ec}

\maketitle

\pub{Received (\today)}{Revised (Day Month Year)}

\begin{abstract}
In this article, we solve the Duffin--Kemmer--Petiau (DKP) equation in the presence of the Woods--Saxon potential barrier and well for spin-one particles. We derive the scattering solution in terms of the Gaussian hypergeometric function ${}_2F_1(a, b; c; z)$, the regularized Gaussian hypergeometric function ${}_2\tilde{F}_1(a, b; c; z)$, and the Gamma function $\Gamma(x)$. Our analysis reveals the presence of transmission resonances. To observe these resonances, we calculate and plot the transmission $T$ and reflection $R$ coefficients for various parameters of the Woods--Saxon potential barrier. Our results are compared with those obtained for the square potential barrier and the cusp potential barrier, which represent limiting cases of the Woods-?Saxon potential barrier. Furthermore, we investigate the bound state solutions, determining the critical turning point $V_{cr}$ to study particle?antiparticle creation and compute the norm $N$. Finally, we also compare our results with those obtained for the square potential well and the cusp potential well, confirming that pair creation occurs in both potential wells.

\end{abstract}

\ccode{PACS Nos.: 03.65.Pm, 03.65.Nk, 03.65.Ge}

\section{Introduction}	


The Duffin--Kemmer--Petiau equation (DKP) plays an important role in modeling systems involving spin--0 and spin--1 particles. The DKP equation has been studied in the context of $(2+1)$ dimensions for a spin--1 particle \cite{hassanabadi:2012} and in the context of $(1+3)$ dimensions for a spin--1 particle with a hyperbolical potential \cite{hassanabadi:2013b}. In recent years, the DKP equation has become a central focus of theoretical studies, especially in the context of one--dimensional spatial potentials \cite{rojas:2024,valladares2023superradiance,langueur2021dkp,sogut:2010,de2010bound,chetouani:2004}.  In particular, some authors have worked with the solution of the DKP equation with the cusp potential barrier and well \cite{rojas:2024}, considering the hyperbolic tangent potential \cite{valladares2023superradiance}. Also, the smooth potential barrier \cite{langueur2021dkp}, the Hulth\'en vector potential \cite{zarrinkamar:2013},  an asymmetric cusp potential \cite{sogut:2010}, and the step potential \cite{chetouani:2004}. In some cases, incorporating position-dependent mass has been considered for the smooth potential barrier \cite{kermezli:2025,merad:2007}, and the square potential well \cite{hammoud:2017}. Also, it has been studied spin-one particle confinement in a two-dimensional-ring potential\cite{CHARGUI2024100818}. Moreover, the DKP theory has been extended to curved spacetimes \cite{lunardi2002interacting,casana2002free} and $\kappa-$Minkowski spacetimes \cite{chargui2020duffin}. Furthermore, the DKP theory has been used to establish a connection between simple theoretical models like classical gauge invariance \cite{lunardi2000remarks}, Galilean covariance \cite{de2000galilean}, and the De Donder?Weyl covariant Hamiltonian \cite{kanatchikov2000duffin}. In general, DKP is a great tool for Theoretical Physics.

The potential that has caught our attention is the Woods--Saxon potential barrier and well \cite{woods:1954}, a well--known model used to describe nuclear forces in the context of spherical symmetry \cite{capak:2016}, the
single--particle motion in a fusion system \cite{diaz2005two}, in heavy--ion physics which is applicable in coupled--channel calculations \cite{hagino2002surface}. In the study of rotating nucleus spectra \cite{molique1997fock}, and generally in the study of confined quantum systems \cite{costa1999study,dakhlaoui2023gaas}. Moreover, it is important to mention that the Woods--Saxon potential has been widely used to describe the nuclear interaction in heavy-ion fusion reactions, especially for synthesizing the superheavy nuclei. Due to their importance in extending the period table of elements and understanding the interplay between nuclear structure and reaction dynamics, machine learning and neural network approaches are considered to determine the best potential parameters in fusion processes \cite{gao2024constraining}.

In the current article, we extend this by addressing the DKP equation in the presence of a Woods--Saxon potential barrier and well. There is a relation between the equation and the potential since both are useful in describing confinement problems of quantum physics in relativistic regions \cite{darroodi:2015} and represent an active area of work. This study examines the scattering and bound states solutions derived from Gaussian hypergeometric functions \cite{villalba2006bound,rojas2006klein,rojas:2005}. A key contribution of this work is the study of transmission resonances, a phenomenon of great interest in quantum scattering theory, along with the analysis of particle-antiparticle bound states. We calculated the transmission coefficient $T$ and the reflection coefficient $R$ for various potential parameters to explore these transmission resonances further.
The transmission resonances obtained for a specific set of parameters in the Woods--Saxon potential barrier are compared with those of the square potential barrier and the cusp potential barrier, which can be regarded as limiting cases of the Woods--Saxon potential.
Furthermore, in the analysis of bound state solutions, we determined the energy $E$ as a function of the depth of the potential well $V_0$ for various values of  $a$  and  $L$. We identified the critical potential  $V_{cr}$, which marks the transition between particle and antiparticle bound states, and used this information to compute the norm  $N$  of each state \cite{greiner:1985}.

A critical potential, denoted as $V_{cr}$, emerges as a pivotal threshold, signifying the potential depth at which relativistic effects become dominant, culminating in the possibility of the creation of a pair of particle--antiparticle \cite{10.1063/1.3047288,greiner2000relativistic}. Mathematically, $V_{cr}$ is defined as the potential depth at which the lowest bound state dives into the negative energy continuum, a phenomenon intrinsically linked to the relativistic nature of the DKP equation \cite{de2010bound}. Physically, exceeding $V_{cr}$ renders the vacuum state unstable, providing sufficient energy to transform virtual particle--antiparticle pairs into real particles \cite{Narlikar1986}. This process mirrors the Klein paradox, in which strong electrostatic potentials induce the creation of electron--positron pairs \cite{Alhaidari_2011}. As the potential depth approaches and exceeds $V_{cr}$, the bound states enter the negative energy continuum, signaling instability, and potential pair creation. Consequently, transmission and reflection coefficients undergo significant changes, resonances arise, and the bound state spectrum transitions from stable states to resonances \cite{10.1142/S0217751X2050195X}.

The norm $N$ of the wave function quantifies the probability density of a quantum state. It was calculated using the energy eigenvalues for different values of  $V_0$  to analyze its behavior and the formation of particle--antiparticle pairs. The norm $N$  changes sign depending on the energy region, indicating a transition in which the bound states merge with the negative energy continuum.

Our particle and antiparticle bound states spectra are compared with those obtained from the square potential well and the cusp potential well. These findings are based on previous studies that solve the DKP equation using the radial form of the Woods?-Saxon potential \cite{boutabia:2016,boutabia2005solution}.

The one--dimensional symmetric Woods--Saxon potential barrier is given by \cite{kennedy:2002},

\begin{equation}
\label{eq001_barrier}
V(x)= {V_0} \left[\dfrac{\Theta (-x)}{1+e^{-a (x+L)}}+\dfrac{\Theta (x)}{1+e^{a (x-L)}}\right],
\end{equation}
where $\Theta(\pm x)$ is the Heaviside step function, given by \cite{kennedy:2002,10.1007/978-3-540-69839-5_93},

\begin{equation}
\Theta(x)=
\left\{
\begin{aligned}
1, \quad x \geq 0,\\
0, \quad x < 0.
\end{aligned}
\right.
\end{equation}

\begin{figure}[th!]
\centering
\includegraphics[scale=0.6]{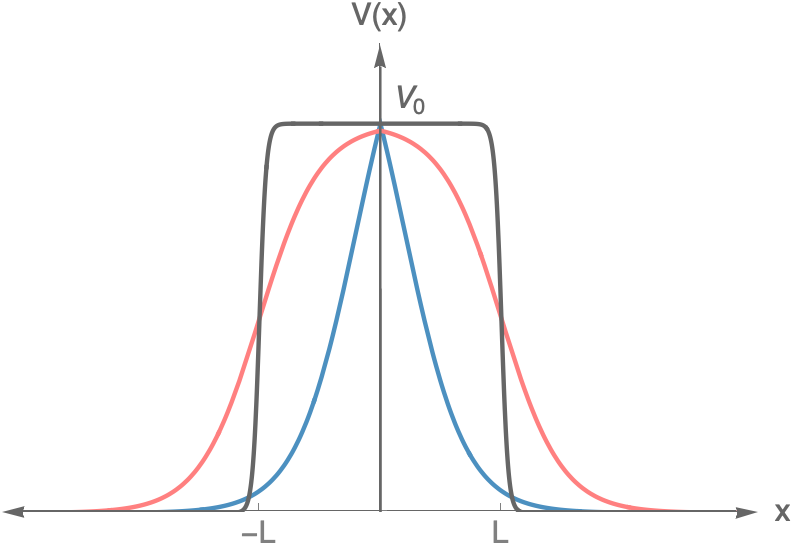}
\caption{Behavior of a Woods--Saxon potential barrier vs the spatial coordinate $x$ for different values of  the parameter parameter $a$. Solid blue light represents $a=2$, $L=0.4$, and $V_0=7.24$, solid pink line represents $a=2$, $L=2$,  and $V_0=5$, finally solid gray line represents $a=18$, $L=2$, and $V_0=5$ . In this figure the parameter $L$ establish the half--width of the potential, and the parameter $a$ its smoothness.}
\label{fig:barrier}
\end{figure}

Consider $a$ as the steepness and $L$ the width parameters of the Woods--Saxon potential barrier, respectively. Moreover, It is important to mention that when $a$ is great enough, the Woods--Saxon potential barrier behaves like a square potential barrier.  The behavior of this barrier is shown in Figure \ref{fig:barrier}.

\begin{figure}[th!]
\centering
\includegraphics[scale=0.6]{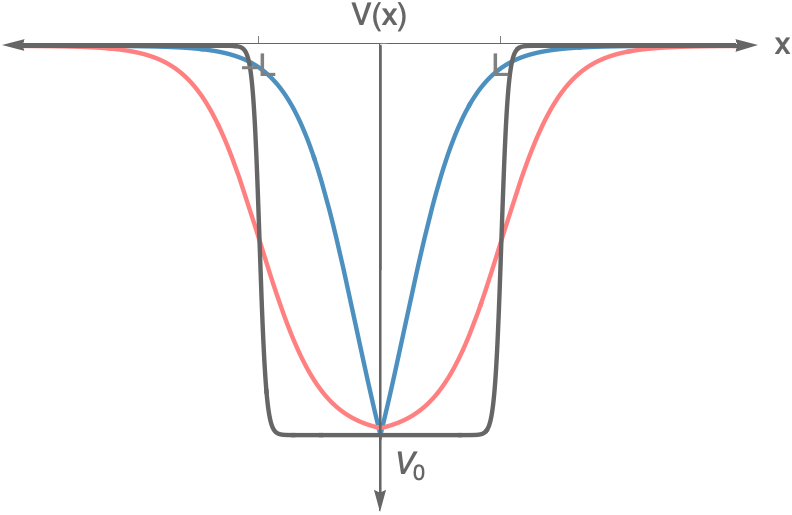}
\caption{Behavior of a Woods--Saxon potential well vs the spatial coordinate $x$ for different values of  the parameter $a$.  Solid blue light represents $a=2$, $L=0.4$, and $V_0=7.24$, solid pink line represents $a=2$, $L=2$,  and $V_0=5$, finally solid gray line represents $a=18$, $L=2$, and $V_0=5$. In this figure the parameter $L$ establish the half--width of the potential, and the parameter $a$ its smoothness.}
\label{fig:well}
\end{figure}

In order to study the bound states solutions, we need study the one--dimensional symmetric Woods--Saxon potential well, which is given by

\begin{equation}
\label{eq002_well}
V(x)= -{V_0} \left[\dfrac{\Theta (-x)}{1+e^{-a (x+L)}}+\dfrac{\Theta (x)}{1+e^{a (x-L)}}\right],
\end{equation}
which is shown in Figure \ref{fig:well}.

The DKP equation interacting with an electromagnetic field is given by (working in natural units $\hbar=c=1$) \cite{langueur2021dkp},

\begin{equation}
\label{eq004}
[i\beta^\mu (\partial_\mu + ieA_\mu) - 1] \Psi(\vec{r}, t) = 0,
\end{equation}
where the matrices $\beta^\mu$ satisfy the DKP algebra \cite{langueur2021dkp}. 
As the potential involved is time--independent, the solution of Eq. \eqref{eq004} can be
obtained by substituting $\Psi(x)=e^{iEt} \phi(x)$. Moreover, the DKP equation can then be rewritten in a $(1+1)$ dimensional representation by \cite{langueur2021dkp},

\begin{equation}
\label{eq005}
\left\{ \beta^0 [E - V(x)] + i\beta^1 \dfrac{\D}{\D x} - 1 \right\} \phi(x) = 0,
\end{equation}
where $\phi(x)$ is the spinor, $E$ is the energy, $V(x)$ represents the one--dimensional spatial potential coupled to the DKP equation. For $(1+1)$ dimensions, we consider the following expressions for $\beta^0$, and $\beta^1$ matrices \cite{rojas:2024,langueur2021dkp}, with the metric tensor $\eta^{\nu\mu}=\textnormal{diag}(1,-1)$,

\begin{eqnarray}
\label{eq006a}
\beta^0 &=& \begin{pmatrix}
0 & 0 & i \\
0 & 0 & 0 \\
-i & 0 & 0
\end{pmatrix}, \\
\label{eq006b}
\beta^1 &=& \begin{pmatrix}
0 & i & 0 \\
i & 0 & 0 \\
0 & 0 & 0
\end{pmatrix}.
\end{eqnarray}

\bigskip
The spinor $\phi(x)$ is presented by the following way \cite{langueur2021dkp},

\begin{equation}
\label{eq007}
\phi(x) = \begin{pmatrix}
\phi_1(x) \\
\phi_2(x) \\
\phi_3(x)
\end{pmatrix},
\end{equation}
where $\phi_1$, $\phi_2$, and $\phi_3$ are the components of the spinor.
Then, expanding Eq. \eqref{eq005} with the definitions of $\beta$ matrices from Eqs. \eqref{eq006a} and \eqref{eq006b}  it reads,

\bigskip
\begin{equation}
\label{equation_exapanded1}
\left\{
\begin{pmatrix}
0 & 0 & i \\
0 & 0 & 0  \\
-i& 0 & 0  
\end{pmatrix} [E-V(x)]+
 i\dfrac{\D}{\D x}
 \begin{pmatrix}
0 & i & 0 \\
i & 0 & 0 \\
0 & 0 & 0
\end{pmatrix}  -\, \mathbb{I} \right\}\begin{pmatrix}
\phi_1(x) \\
\phi_2(x)\\
\phi_3 (x) \end{pmatrix} = 0
.
\end{equation}

\bigskip
After some computation, the matrices become.

\bigskip
\begin{equation}
\label{equation_expanded_2}
\left\{
\begin{pmatrix}
0 & 0 & i\, [E-V(x)] \\
0 & 0 & 0  \\
-i\, [E-V(x)]& 0 & 0  
\end{pmatrix}+
 \begin{pmatrix}
0 & -\dfrac{\D}{\D x} & 0 \\
-\dfrac{\D}{\D x} & 0 & 0 \\
0 & 0 & 0
\end{pmatrix}  -\, \mathbb{I}  \right\}\begin{pmatrix}
\phi_1(x) \\
\phi_2(x)\\
\phi_3 (x)  
\end{pmatrix}=0.
\end{equation}

\bigskip\bigskip
Performing the addition of the matrices,

\bigskip
\begin{equation}
\label{equation_expanded_3}
\left\{
\begin{pmatrix}
0 & -\dfrac{\D}{\D x} & i\, [E-V(x)] \\
-\dfrac{\D}{\D x} & 0 & 0  \\
-i\, [E-V(x)]& 0 & 0  
\end{pmatrix}  
-\, \mathbb{I} \right\}\begin{pmatrix}
\phi_1(x) \\
\phi_2(x)\\
\phi_3 (x)  
\end{pmatrix}= 0.
\end{equation}

\bigskip
Operating, this equation follows the next shape,

\begin{equation}
\label{equation_expanded_4}
\left\{
\begin{pmatrix}
0 & -\dfrac{\D}{\D x} & i\, [E-V(x)] \\
-\dfrac{\D}{\D x} & 0 & 0  \\
-i\, [E-V(x)]& 0 & 0  

\end{pmatrix}\begin{pmatrix}
\phi_1(x) \\
\phi_2(x)\\
\phi_3(x)  
\end{pmatrix}
-\begin{pmatrix}
\phi_1(x) \\
\phi_2(x)\\
\phi_3(x)  
\end{pmatrix} \right\}= 0.
\end{equation}

\bigskip
Then, this becomes in the next sum that should be decomposed into an equation system,

\bigskip
\begin{equation}
\label{equation_expanded_final}
\left\{
\begin{pmatrix}
0 & -\dfrac{\D \phi_2(x)}{\D x} & i\,[E-V(x)]\phi_3 (x) \\
-\dfrac{\D \phi_1(x)}{\D x} & 0 & 0  \\
-i\,[E-V(x)]\phi_1(x)& 0 & 0  

\end{pmatrix}
-\begin{pmatrix}
\phi_1(x) \\
\phi_2(x)\\
\phi_3(x)  
\end{pmatrix} \right\}= 0.
\end{equation} 

\bigskip
Finally, from the previous expression, it is possible to have the row representation of the system of equations as follows,

\bigskip
\begin{equation}
\label{Dkpequation_system}
\left\{
\begin{aligned}
&\left\{\dfrac{\D^2}{\D x^2}  + [E - V(x)]^2 - 1\right\}\, \phi_1(x) = 0,\\
&\phi_2(x) = -\dfrac{\D}{\D x} \phi_1(x), \\
&\phi_3(x) = -i \, [E - V(x)]\, \phi_1(x),
\end{aligned}
\right.
\end{equation}
where $\phi_1(x)$ depends on the solution of the Klein--Gordon equation. Generally, expressions \eqref{Dkpequation_system} are the main problem to solve for the potential barrier and well.

In recent years, several analytical and numerical methods have been proposed and employed to solve the Duffin--Kemmer--Petiau (DKP) equation under various potential wells. Among these methods are the factorization approach, supersymmetric quantum mechanics (SUSY--QM), as suggested in \cite{okninski2011supersymmetric}, the Nikiforov--Uvarov method \cite{hassanabadi2013exact}, and the asymptotic iteration method (AIM) \cite{boztosun2006asymptotic}. One particularly elegant and powerful technique arises when the DKP equation is of the form given by Eq. \eqref{Dkpequation_system}, is reduced to a second--order differential equation whose solutions can be expressed in terms of special functions, most notably, the hypergeometric function, as demonstrated in \cite{valladares2023superradiance,rojas:2024}.

In this work, we focus on the resolution of the DKP equation in the expression provided by Eq. \eqref{Dkpequation_system} using a method that reduces it to a hypergeometric--type differential equation. After an appropriate change of variables, we obtain a second--order differential equation of the form,

\begin{equation}
z(1 - z)\, \dfrac{\D^2 y(z)}{\D z^2} + [c - (a + b + 1)\,z] \,\dfrac{\D y(z)}{\D z} - a\,b\, y(z) = 0,
\end{equation}
which is the standard form of the Gaussian hypergeometric equation \cite{abramowitz:1965}. 

The parameters $a, b, c$ are determined from the coefficients of the DKP equation after transformation, and the solution is written in terms of the Gaussean hypergeometric function,

\begin{equation}
y(z) = {}_2F_1(a, b; c; z),
\end{equation}
where ${}_2F_1(a, b; c; z)$ denotes the Gaussian hypergeometric function \cite{abramowitz:1965}.

The Gaussian hypergeometric function ${}_2F_1(a, b; c; z)$ is defined by the following power series expansion \cite{abramowitz:1965},

\begin{equation}
{}_2F_1(a, b; c; z) = \sum_{n=0}^{\infty} \dfrac{(a)_n (b)_n}{(c)_n} \dfrac{z^n}{n!},
\end{equation}

where:
\begin{itemize}

\item The expression $(a)_n$, $(b)_n$, and $(c)_n$ are the Pochhammer symbols \cite{lozier2003nist} (also known as rising factorials ), defined by,

\begin{equation}
(q)_n = q (q+1) (q+2) \cdots (q+n-1) = \dfrac{\Gamma(q+n)}{\Gamma(q)}, \quad (q)_0 = 1,
\end{equation}

\item $z$ is the independent variable (the series converges for $|z| < 1$ when $c \notin \mathbb{Z}{\leq 0}$).

\item $\Gamma(z)$ is the gamma function, which generalizes the factorial to complex numbers.
\end{itemize}

This method not only provides exact solutions for the spinor components, but also allows the computation of physically relevant quantities such as reflection and transmission coefficients, and facilitates the study of resonance phenomena in scattering scenarios \cite{rojas:2024,valladares2023superradiance}.

This paper is organized as follows. In Section 2, we solve the DKP equation for the Woods--Saxon potential barrier, calculate the transmission coefficient $T$ and the reflection coefficient $R$, and observe the transmission resonances for several parameters of the barrier. In Section 3, we study the bound states for the Woods--Saxon potential well by solving the DKP equation, finding the particle--antiparticle bound states, and calculating the norm for several cases. In Section 2 and Section 3 we compare our results for the Woods--Saxon potential barrier and well considering a big value for $a$ with the results for a square potential barrier and well, also for different sets of parameters we compare our results for the Woods--Saxon potential barrier and well with those obtained for a cusp potential barrier and well.  Section 4 is devoted to the discussion of the conclusions of our work. Finally,  Section 5 is devoted to express the acknowledgments of the authors.

\section{Transmission Resonances}

\subsection{For $x<0$}

In order to compute the transmission resonances, we solve Eqs. \eqref{Dkpequation_system} with the Woods--Saxon potential barrier, which is given by Eq. \eqref{eq001_barrier}, then it is necessary to find the first component of Eq. \eqref{Dkpequation_system} to find the first spinors component $\phi_1(x)$ (where in this expression $\phi_{1L}(x)$, $L$ is related to the left hand side),

\begin{equation}
\label{eq015}
\dfrac{\D^2\phi_{1L}(x)}{\D x^2} + \left\{\left[E - \dfrac{V_0}{1+e^{-a(x+L)}}\right]^2 - 1 \right\} \phi_{1L}(x) = 0.
\end{equation}

The next step is to change the variable $x$ for $y = -e^{-a(x+L)}$, which leads to Eq. (\ref{eq015}) in the form,

\begin{equation}
\label{eq016}
a^{2}y\dfrac{\D}{\D y}\left[y\dfrac{\D \phi_{1L}(y)}{\D y}\right] + \left\{\left[E - \dfrac{V_0}{1-y}\right]^2 - 1 \right\} \phi_{1L}(y) = 0.
\end{equation}

Putting $\phi_{1L}(y) = y^{\mu}(1-y)^{-\lambda_1} h(y)$, Eq. \eqref{eq016} becomes such suggest and took the form of a hypergeometric differential equation \cite{abramowitz:1965},

\begin{equation}
\label{eq017}
y (1-y) \,h'' + [(1+2\mu)-(2\mu -2\lambda_1 + 1)y] \,h' - (\mu -\lambda_1 +\nu)(\mu - \lambda_1 - \nu) \,h = 0,
\end{equation}
where prime means derivative respect to $y$, and  parameters $\nu, k, \mu, \lambda, \lambda_1$ becomes,

\begin{eqnarray}
\label{eq018_nu}
\nu &=& \dfrac{ik}{a},\\
k &=& \sqrt{E^2 - 1},\\
\mu &=& \dfrac{\sqrt{1 - (E - V_0)^2}}{a},  \\
\label{eq021_lambda}
\lambda &=& \dfrac{\sqrt{a^2 - 4V_0^2}}{2a},\\
\lambda_1 &=& -\dfrac{1}{2} + \lambda. 
\end{eqnarray}

The general solution of Eq. (\ref{eq017}) is expressed in terms of Gaussian hypergeometric functions  such as \cite{abramowitz:1965},

\begin{eqnarray}
\nonumber
h(y) &=& D_{1}\, {}_2F_{1}(\mu -\nu -\lambda_1, \mu + \nu -\lambda_1; 1 + 2\mu; y) \\
& +&  D_2 \, y^{-2\mu} {}_2F_{1}(-\mu -\nu -\lambda_1, -\mu + \nu - \lambda_1; 1 - 2\mu; y),
\label{eq023}
\end{eqnarray}
so on $\phi_{1L}(y)$ becomes,

\begin{eqnarray}
\nonumber
\phi_{1L}(y) &=& D_{1}\, y^{\mu}(1-y)^{-\lambda_1} {}_2F_{1}(\mu -\nu -\lambda_1, \mu + \nu -\lambda_1; 1 + 2\mu; y) \\
 &+& D_2 \,y^{-\mu}(1-y)^{-\lambda_1} {}_2F_{1}(-\mu -\nu -\lambda_1, -\mu + \nu - \lambda_1; 1 - 2\mu; y).
\label{eq024}
\end{eqnarray}

\bigskip
Realizing that present solution Eq. (\ref{eq024}) is made up of the incident and the reflected part of the wave function. Furthermore, the objective of the set is to solve the system of equations presented in Eq. (\ref{Dkpequation_system}). Consequently, we will examine each case individually using the DKP Algebra. This ensures a comprehensive understanding of the mathematical framework.

\subsubsection{Incident part}

The solution of the system of equations Eq. (\ref{Dkpequation_system}) turning back the $y(x)$ change of variable so the incident part is presented to us such, 

\begin{eqnarray}
\nonumber
\phi_{{1L}_{\text{inc}}}(x) 
&=& D_1\, \left[-e^{-a (x+L)}\right]^{\mu}\left[1+e^{-a (x+L)} \right]^{-\lambda_1}  \\
&\times& {}_2F_1\left[-\lambda_1 + \mu - \nu, -\lambda_1 + \mu + \nu; 1+2\mu ; -e^{-a (x+L)}\right],\\
\nonumber
\phi_{{2L}_{\text{inc}}}(x)  &=& -a\, D_1\, \lambda_1 \, e^{-a (x+L)} \left[-e^{-a (x+L)}\right]^{\mu}  \left[1+e^{-a (x+L)}\right]^{-1-\lambda_1} \\
\nonumber
&\times& {}_2F_1\left[-\lambda_1+\mu-\nu, -\lambda_1+\mu+\nu; 1+2\mu; -e^{-a (x+L)}\right] \\
\nonumber
&-& a\, D_1\, \mu\, e^{-a (x+L)} \left[-e^{-a (x+L)}\right]^{-1+\mu} \left[1+e^{-a (x+L)}\right]^{-\lambda_1} \\
\nonumber
&\times& {}_2F_1\left[-\lambda_1+\mu-\nu, -\lambda_1+\mu+\nu; 2\mu+1; -e^{-a (x+L)}\right] \\
\nonumber
&-& \dfrac{a\, D_1}{\left(1+2\mu\right)}  (-\lambda_1+\mu-\nu)(-\lambda_1+\mu+\nu) e^{-a (x+L)} \left[-e^{-a (x+L)}\right]^{\mu} \left[1+e^{-a (x+L)}\right]^{-\lambda_1}\\
&\times&  {}_2F_1\left[-\lambda_1+\mu-\nu+1, -\lambda_1+\mu+\nu+1; 2\, (1+\mu); -e^{-a (x+L)}\right],\\
\nonumber
\phi_{{3L}_{\text{inc}}}(x)  &=& -i\, D_1 \, \left[-e^{-a (x+L)}\right]^{\mu} \left[1+e^{-a (x+L)} \right]^{-\lambda_1} \left[E-\dfrac{V_0}{1+e^{-a (x+L)}}\right]\\
\nonumber
&\times& {}_2F_1\left[-\lambda_1+\mu-\nu, -\lambda_1+\mu+\nu; 1
+2\mu; -e^{-a (x+L)}\right].\\\\
\label{eq025}
\nonumber
\end{eqnarray}

Then, all this three elements becomes each component of the next spinors, furthermore the long distance in each can not be fixed in a single page, in spite of it they could be read in a straightforward shape as next,

\begin{equation}
\label{vec_spinor_inc}
\phi_{\text{inc}}(x) = 
\begin{pmatrix}
\phi_{{1L}_{\text{inc}}}(x) \\
\phi_{{2L}_{\text{inc}}}(x) \\
\phi_{{3L}_{\text{inc}}}(x) 
\end{pmatrix}.
\end{equation}

\subsubsection{Reflected part}
After has been solved the side of the system for the incident part it is needed to find the spinor for the reflected side of the wave. Moreover, the solution of the system of Eq. \eqref{Dkpequation_system} turning back the $y(x)$ change of variable to the reflected part in such a way,

\begin{eqnarray}
\nonumber
\phi_{{1L}_{\text{ref}}}(x) 
&=& D_2\, \left[-e^{-a (x+L)}\right]^{-\mu} \left[1+e^{-a (x+L)}\right]^{-\lambda_1} \ \, \\
&\times& {}_2F_1\left[-\lambda_1 - \mu - \nu, -\lambda_1 - \mu + \nu;  1 - 2\mu ; -e^{-a (x+L)}\right],\\
\nonumber
\phi_{{2L}_{\text{ref}}}(x) 
&=& -a\, D_2\, \lambda_1\,  e^{-a (x+L)} \left[-e^{-a (x+L)}\right]^{-\mu} \left[1+e^{-a (x+L)}\right]^{-1-\lambda_1}  \\
\nonumber
&\times&  {}_2F_1\left[-\lambda_1 - \mu - \nu, -\lambda_1 - \mu + \nu;  1 - 2\mu ; -e^{-a (x+L)}\right] \\
\nonumber
& +& a\, D_2\, \mu \,e^{-a (x+L)} \left[-e^{-a (x+L)}\right]^{-1-\mu} \left[1+e^{-a (x+L)}\right]^{-\lambda_1}  \\
\nonumber
&\times& {}_2F_1\left[-\lambda_1 - \mu - \nu, -\lambda_1 - \mu + \nu;  1 - 2\mu ; -e^{-a (x+L)}\right] \\
\nonumber
&-& \dfrac{a\, D_2}{\left(1-2\mu\right)}  (-\lambda_1-\mu-\nu)(-\lambda_1-\mu+\nu) e^{-a (x+L)} \left[-e^{-a (x+L)}\right]^{-\mu}   \left[1+e^{-a (x+L)}\right]^{-\lambda_1} \\
&\times& {}_2F_1\left[1 -\lambda_1 - \mu - \nu, 1 -\lambda_1 - \mu + \nu;  2\,(1 - \mu) ; -e^{-a (x+L)}\right],\\ \nonumber
\phi_{{3L}_{\text{ref}}}(x) 
&=& -i\, D_2 \left[-e^{-a (x+L)}\right]^{-\mu } \left[1+e^{-a (x+L)}\right]^{-\lambda_1}  \left[E-\dfrac{V_0}{1+e^{-a (x+L)}}\right] \\
&\times& {}_2F_1\left[-\lambda_1 - \mu - \nu, -\lambda_1 - \mu + \nu;  1 - 2\mu ; -e^{-a (x+L)}\right] .\\ \nonumber
\label{eq027}
\label{eq028}
\label{eq029}
\end{eqnarray}

Bound in one vector, this is where there is no sense coupling each value of the whole vector,

\begin{equation}
\label{vec_spinor_reflec}
\phi_{\text{ref}}(x) = 
\begin{pmatrix}
\phi_{{1L}_{\text{ref}}}(x) \\
\phi_{{2L}_{\text{ref}}}(x) \\
\phi_{{3L}_{\text{ref}}}(x) 
\end{pmatrix}.
\end{equation}

\subsection{For $x>0$:}

Now we consider the solution for $x>0$. In this case,the differential equation to solve is,
 
\begin{equation}
\label{eq031}
\dfrac{\D^2\phi_{1R}(x)}{\D x^2} + \left\{\left[E - \dfrac{V_0}{1+e^{a(x-L)}}\right]^2 - 1 \right\} \phi_{1R}(x) = 0.
\end{equation}

In order to operate this equation, the next change of variable is used: $z^{-1} = 1 + e^{a(x-L)}$, which leads to Eq. (\ref{eq031}) to the following differential form,

\begin{equation}
\label{eq032}
a^{2}z(1-z) \dfrac{\D}{\D z}\left[z(1-z)\dfrac{\D\phi_{1R}(z)}{\D z}\right] + \left[(E-V_0 \,z)^{2}-1\right]\phi_{1R}(z) =0.
\end{equation}

Following the line, setting the dependence of $\phi_{R}(z)$ such that $\phi_{1R}(z) = z^{-\nu}(1-z)^{-\mu} g(z)$, Eq.  (\ref{eq032}) becomes the hypergeometric equation,

\begin{equation}
\label{eq033}
z(1-z) \,g'' + \left[ (1-2\nu) - 2(1-\nu-\mu)z\right] \,g'-\left(\dfrac{1}{2}-\nu-\mu -\lambda\right)\left(\dfrac{1}{2}-\nu -\mu +\lambda\right) \,g = 0, 
\end{equation}
where the primes denote derivatives respect to $z$. Furthermore, the solution to Eq. (\ref{eq033}) is,

\begin{eqnarray}
\nonumber
g(z) &=& d_{1} {}_2F_{1}\left(\dfrac{1}{2}-\nu-\mu -\lambda,\dfrac{1}{2}-\nu -\mu +\lambda; 1 - 2\nu; z\right) \\
& +&  d_2 \, z^{2\nu} {}_2F_{1}\left(\dfrac{1}{2}+\nu-\mu -\lambda,\dfrac{1}{2}+\nu -\mu +\lambda; 1 + 2\nu; z\right),
\label{eq034}
\end{eqnarray}
so the general solution for Eq. \eqref{eq031} becomes  (where in  $\phi_{1R}(z)$, $R$ is related to the right hand side),

\begin{eqnarray}
\nonumber
\phi_{1R}(z) &=& d_{1}\, z^{-\nu} (1-z)^{-\mu} {}_2F_{1}\left(\dfrac{1}{2}-\nu-\mu -\lambda,\dfrac{1}{2}-\nu -\mu +\lambda; 1 - 2\nu; z\right) \\
&+& d_2\, z^{\nu}(1-z)^{-\mu} {}_2F_{1}\left(\dfrac{1}{2}+\nu-\mu -\lambda,\dfrac{1}{2}+\nu -\mu +\lambda; 1 + 2\nu; z\right).
\label{eq035}
\end{eqnarray}

\bigskip
Next, because the transmitted side of the wave is just composed of the equation $d_{1}$, we have $d_{2} = 0$ in the equation above. Moreover, Eq. \eqref{eq035} becomes,

\bigskip
\begin{equation}
\label{eq036}
\phi_{1R}(z) = d_{1}\,z^{-\nu} (1-z)^{-\mu} {}_2F_{1}\left(\dfrac{1}{2}-\nu-\mu -\lambda,\dfrac{1}{2}-\nu -\mu +\lambda; 1 - 2\nu; z\right). \textsuperscript{b}
\end{equation}

\subsubsection{Transmitted part}

Turning back the change of variable for $z(x)$  the components of the transmitted vector becomes 

\begin{eqnarray}
\nonumber
\phi_{{1R}_{\text{trans}}}(x) 
&=& d_1\,\left[1+e^{a (x-L)}\right]^{\nu}\left[1-\dfrac{1}{1+e^{a (x-L)}}\right]^{-\mu} \\
&\times& {}_2F_1\left[\dfrac{1}{2}-\lambda -\mu -\nu ,\dfrac{1}{2}+\lambda -\mu -\nu ;1-2 \nu ;\dfrac{1}{1+e^{a (x-L)}}\right],
\end{eqnarray}

\begin{eqnarray}
\label{eq037}
\nonumber
\phi_{{2R}_{\text{trans}}}(x) &=& a\, d_1\,  \mu \, e^{a (x-L)}   \left[1+e^{a (x-L)}\right]^{-2+\nu}\left[1-\dfrac{1}{1+e^{a (x-L)}}\right]^{-1-\mu }\\
\nonumber
&\times&  {}_2F_1\left[\dfrac{1}{2}-\lambda -\mu -\nu ,\dfrac{1}{2}+\lambda -\mu -\nu ;1-2 \nu ;\dfrac{1}{1+e^{a (x-L)}}\right] \\
\nonumber
&-& a\, \nu\, d_1\, e^{a (x-L)} \left[1+e^{a(x-L)}\right]^{-1+\nu} \left[1-\dfrac{1}{1+e^{a (x-L)}}\right]^{-\mu } \\
\nonumber
&\times& {}_2F_1\left[\dfrac{1}{2}-\lambda -\mu -\nu ,\dfrac{1}{2}+\lambda -\mu -\nu ;1-2 \nu ;\dfrac{1}{1+e^{a (x-L)}}\right] \\
\nonumber
&+& \dfrac{a \, d_1}{\left(1-2 \nu\right)} \Bigg\{ \left(\dfrac{1}{2}-\lambda -\mu -\nu\right) \left(\dfrac{1}{2}+\lambda -\mu -\nu \right) \\
\nonumber
&\times& e^{a (x-L)} \left[1+e^{a (x-L)}\right]^{-2+\nu}\left[1-\dfrac{1}{1+e^{a (x-L)}}\right]^{-\mu } \\
&\times& {}_2F_1\left[\dfrac{3}{2}-\lambda -\mu -\nu ,\dfrac{3}{2}+\lambda -\mu -\nu; 2 \, (1- \nu) ;\dfrac{1}{1+e^{a (x-L)}}\right] \Bigg\},\\
\nonumber
\phi_{{3R}_{\text{trans}}}(x) 
&=& -i\, d_1 \left[1+e^{a (x-L)}\right]^{\nu } \left[1-\dfrac{1}{1+e^{a (x-L)}}\right]^{-\mu }   \left[E-\dfrac{V_0}{1+e^{a (x-L)}}\right] \nonumber\\
&\times& {}_2F_1\left[\dfrac{1}{2}-\lambda - \mu - \nu, \dfrac{1}{2}+\lambda - \mu - \nu;  1 - 2\nu ; \dfrac{1}{1+e^{a (x-L)}}\right].
\label{eq038}
\end{eqnarray}

\bigskip
Then, the vector composed looks like,

\bigskip
\begin{equation}
\label{vec_spinortrans}
\phi_{\text{trans}}(x) = 
\begin{pmatrix}
\phi_{{1R}_{\text{trans}}}(x) \\
\phi_{{2R}_{\text{trans}}}(x) \\
\phi_{{3R}_{\text{trans}}}(x) 
\end{pmatrix},
\end{equation}
where this new vector is the spinor of the transmitted wave.

After all these coupled vectors have been formed, we need to analyze the asymptotic behavior for each one.

\subsection{Asymptotic behavior for  $\phi_{\textnormal{inc}}(x), \phi_{\textnormal{ref}}(x),\phi_{\textnormal{trans}}(x)$.}

In order to find the scattering states for the Woods--Saxon potential barrier using the DKP spinors, the asymptotic behavior must be analyzed. Moreover, Gaussian hypergeometric functions could be written in terms of Gamma functions $\Gamma(x)$ in the following way \cite{abramowitz:1965},

\begin{equation}
\label{eq041}
{}_2F_{1}(a, b; c; z) = \dfrac{\Gamma(c) \Gamma(b - a)}{\Gamma(b) \Gamma(c - a)} (-z)^{-a} + \dfrac{\Gamma(c) \Gamma(a - b)}{\Gamma(a) \Gamma(c - b)} (-z)^{-b}.
\end{equation}

This means that for the $y$ tendency at infinity, it becomes $x \to -\infty$ then $(-y)^{\pm\nu} \to  e^{\pm ik(x+L)}$, so using these two mathematical arguments the next components of the spinor $\phi_{1L}(y)$ element, Eq. \eqref{Dkpequation_system}, should be written as

\begin{equation}
\label{eq042}
\phi_{1L}(x) = A\, e^{ik(x+L)} + B\, e^{-ik(x+L)}.
\end{equation}

Then, some computing arguments develop the $A$ and $B$ scalars in terms of the $D_1$ and $D_2$ scalars as follows,

\begin{eqnarray}
\label{eq043}
\nonumber
A &=& D_1\,\dfrac{\Gamma(1+2\mu) \Gamma(-2\nu)}{\Gamma(\mu-\nu -\lambda_1) \Gamma(1+\mu-\nu+\lambda_1)} (-1)^{\mu} \\
&+& D_2\,\dfrac{\Gamma(1-2\mu) \Gamma(-2\nu)}{\Gamma(\mu -\nu -\lambda_1) \Gamma(1-\mu -\nu +\lambda_1)} (-1)^{-\mu},\\
\nonumber\\
\label{eq044}
\nonumber
B &=& D_1\,\dfrac{\Gamma(1+2\mu) \Gamma(2\nu)}{\Gamma(\mu+\nu -\lambda_1) \Gamma(1+\mu+\nu+\lambda_1)} (-1)^{\mu} \\
&+& D_2\,\dfrac{\Gamma(1-2\mu) \Gamma(2\nu)}{\Gamma(-\mu +\nu -\lambda_1) \Gamma(1-\mu +\nu +\lambda_1)} (-1)^{-\mu}. 
\end{eqnarray}

\bigskip
Taking into account the added constants $A$ and $B$, then the spinors for each side will have a different equation shape, such as

\begin{equation}
\label{spinor4}
\phi_{\text{inc}}(x)= A\, e^{ik(x+L)}  
\begin{pmatrix}
1\\
-ik \\
-iE
\end{pmatrix}.
\end{equation}

The result looks quite similar to a wave equation moving in a direction. 
For the reflected wave, the spinors looks like:

\begin{equation}
\label{spinor5}
\phi_{\text{ref}}(x)= B\, e^{-ik(x+L)}  
\begin{pmatrix}
1\\
ik \\
-iE
\end{pmatrix}.
\end{equation}

In order to compute the transmitted side spinor, then the same mathematical computing should be applied with $x\to \infty$,

\begin{equation}
\label{spinor6}
\phi_{\text{trans}}(x)= d_1\, e^{ik(x-L)}  
\begin{pmatrix}
1\\
-ik \\
-iE
\end{pmatrix}.
\end{equation}

\bigskip
With the aim information of the new spinors, it can be developed the transmission $T$ and the reflection $R$ coefficients.

\bigskip
The four--vector current density $j^\mu$ for the DKP equation is given by \cite{cardoso2010effects} \cite{cardoso2010nonminimal}

\begin{equation}
j^{\mu} = \bar{\phi} \beta^{\mu} \phi,
\end{equation}
where   $\bar{\phi}$ is the adjoint conjugate of the spinors associated. 

Using this expression, one can determine the transmission  and  reflection coefficients, \(T\) and \(R\),  respectively. To compute the transmission \(T\) and reflection \(R\) coefficients, we impose the continuity of the wave function at \(x = 0\), furthermore using the spinors of Eqs (\eqref{spinor4},\eqref{spinor5} and \eqref{spinor6}) the coefficients becomes,
\cite{sogut:2010}.

\begin{eqnarray}
\label{eq045b}
T &=& \left|\dfrac{j_\textnormal{trans}}{j_\textnormal{inc}}\right|^{2} = \left|\dfrac{d_1}{A}\right|^{2},\\
\label{eq045a}
R &=& \left|\dfrac{j_\textnormal{ref}}{j_\textnormal{inc}}\right|^{2} = \left|\dfrac{B}{A}\right|^{2},
\end{eqnarray}
which satisfy the unitary relation,

\begin{equation}
R + T = 1.
\end{equation}

\subsection{Transmission and Reflection coefficients}

The next step is to plot the transmission coefficient $T$ and the reflection coefficient $R$, but to do this, there it is necessary to compute the identity of the constants $D_1$ and $D_2$. At this point, we need to develop a new system of equations with the spinor to know the identity of the linear dependence of the coefficients. Furthermore, identification of the entire general solution for the DKP equation evaluating its continuity at $x=0$,

\begin{equation}
\phi_\textnormal{inc} (x) \Big|_{x = 0} +  \phi_\textnormal{ref} (x) \Big|_{x = 0} =  \phi_\textnormal{trans} (x) \Big|_{x = 0}.
\label{eq047continuity}
\end{equation}

After this each component has a tendency to zero it means that the sym of the vector is done element by element, this implies that $\phi_{1L\text{inc}}$ + $\phi_{1R\text{trans}}$ = $\phi_{1L\text{ref}}$ following this argument and axiom for the vector components summation the others equations can be read as, $\phi_{2L\text{inc}}$ +  $\phi_{2L\text{ref}}$= $\phi_{2R\text{trans}}$ and then the same for the third component, $\phi_{3L\text{inc}}$ +  $\phi_{3L\text{ref}}$= $\phi_{3R\text{trans}}$ 
Using the condition \eqref{eq047continuity} the linear system becomes,

\begin{eqnarray}
\label{eq052}
\nonumber
& & D_1 \left(-e^{-a L}\right)^{-\mu}\left(1+e^{-a L}\right)^{-\lambda_1}   {}_2F_1\left(-\lambda_1+\mu -\nu, -\lambda_1+\mu +\nu; 1-2\mu ; -e^{-a L}\right) \\
\nonumber
&+&D_2\left(-e^{-a L}\right)^{\mu} \left(1+e^{-a L}\right)^{-\lambda_1}   {}_2F_1\left(-\lambda_1-\mu -\nu, -\lambda_1+\mu +\nu; 1-2\mu; -e^{-a L}\right) \\
\nonumber
&=& d_1 \left(1+e^{-a L}\right)^{\nu} \left(1-\dfrac{1}{1+e^{-a L}}\right)^{-\mu}    {}_2F_1\left(\dfrac{1}{2}-\lambda -\mu -\nu, \dfrac{1}{2}+\lambda -\mu -\nu; 1-2\nu; \dfrac{1}{1+e^{-a L}}\right),\\
\nonumber\\\\
\label{eq053}
\label{eq2}
\nonumber
&+& D_2 \left(-e^{-a L}\right)^{-\mu } \left(1+e^{-a L}\right)^{-\lambda_1}  {}_2F_1\left(-\lambda_1-\mu -\nu ,-\lambda_1-\mu +\nu ;1-2 \mu ;-e^{-a L}\right)\\
\nonumber
&+& D_1 \left(-e^{-a L}\right)^{\mu } \left(1+e^{-a L}\right)^{-\lambda_1}  {}_2F_1\left(-\lambda_1+\mu -\nu ,-\lambda_1+\mu +\nu ;1+2 \mu;-e^{-a L}\right)\\
&=& d_1 \left(1+e^{-a L}\right)^{\nu }\left(1-\dfrac{1}{1+e^{-a L}}\right)^{-\mu } 
{}_2F_1\left(\dfrac{1}{2}-\lambda -\mu -\nu,\dfrac{1}{2}+\lambda -\mu -\nu;1-2 \nu ;\dfrac{1}{1+e^{-a L}}\right),
\nonumber\\\\
\label{eq3}
\nonumber
&-& a\,  D_1\, \lambda_1\, e^{-a L} \left(-e^{-a L}\right)^{\mu} \left(1+e^{-a L}\right)^{-1-\lambda_1} 
h_1 \\
\nonumber
&+& a\, D_1\, \mu\, e^{-a L}\left(-e^{-a L}\right)^{-1+\mu} \left(1+e^{-a L}\right)^{-\lambda_1} 
h_1 \\
\nonumber
&-& \left(\dfrac{1}{1+2 \mu} \right) a\, D_1\, e^{-a L} \left(-e^{-a L}\right)^{-\mu} \left(1+e^{-a L}\right)^{-\lambda_1} 
(-\lambda_1+\mu -\nu) (-\lambda_1+\mu +\nu)\,
h_2 \\
\nonumber
&-& a\, D_2\, \lambda_1\, e^{-a L} \left(-e^{-a L}\right)^{\mu} \left(1+e^{-a L}\right)^{-1-\lambda_1} 
h_3 \\
\nonumber
&+& a\,  D_2\, \mu\, e^{-a L} \left(-e^{-a L}\right)^{-1+\mu} \left(1+e^{-a L}\right)^{-\lambda_1} 
h_3 \\
\nonumber
&-& \left(\dfrac{1}{1-2 \mu}\right) a\, D_2\, e^{-a L} \left(-e^{-a L}\right)^{-\mu} \left(1+e^{-a L}\right)^{-\lambda_1} 
(-\lambda_1-\mu -\nu) (-\lambda_1-\mu +\nu)\,
h_4 \\
\nonumber 
& = & a\, d_1\, \mu\, e^{-a L} \left(1+e^{-a L}\right)^{-2+\nu}\left(1-\dfrac{1}{1+e^{-a L}}\right)^{-1-\mu}
h_5 \\
\nonumber
&-& a\, d_1\, \nu\, e^{-a L} \left(1+e^{-a L}\right)^{-1+\nu} \left(1-\dfrac{1}{1+e^{-a L}}\right)^{-\mu}
h_5 \\
\nonumber
& +& \left(\dfrac{1}{1-2 \nu}\right) \Bigg\{ a\, d_1\, e^{-a L} \left(1+e^{-a L}\right)^{-2+\nu}\left(1-\dfrac{1}{1+e^{-a L}}\right)^{-\mu}  \\
\nonumber
&\times& \left(\dfrac{1}{2}-\lambda -\mu -\nu \right) \left(\dfrac{1}{2}+\lambda -\mu -\nu \right)\,
h_6 \Bigg\}.\\
\end{eqnarray}

Due to the long of the third component $\phi_{3L\text{inc}} + \phi_{3R\text{trans}} = \phi_{3L\text{ref}}$,  and in order to make Eq. \eqref{eq3} more readable, we  use Table \ref{tab:hypergeometric_functions} to write this long equation in a compact format, where the functions $h_i$ are defined in terms of the Gaussian hypergeometric function and the regularized Gaussian hypergeometric function, which is given by,

\begin{equation}
{}_2\tilde{F}_1\left(a,b;c;z\right)=\dfrac{{}_2{F}_1\left(a,b;c;z\right)}{\Gamma(c)}.
\end{equation}

\begin{table}[th!]
\centering
\caption{Definitions of functions $h_i$ used in our calculations.}
\begin{tabular}{c l}
\toprule
Function & Definition \\ \midrule
$h_1$    & ${}_2F_1\left(-\lambda_1 + \mu - \nu, -\lambda_1 + \mu + \nu; 1+2\mu; -e^{-a L}\right)$ \\
$h_2$    & ${}_2F_1\left(1-\lambda_1 + \mu - \nu,1 -\lambda_1 + \mu + \nu; 2 \,(1 + \mu); -e^{-a L}\right)$ \\
$h_3$    & ${}_2F_1\left(-\lambda_1 - \mu - \nu, -\lambda_1 - \mu + \nu; 1 - 2\mu; -e^{-a L}\right)$ \\
$h_4$    & ${}_2F_1\left(1-\lambda_1 - \mu - \nu, 1-\lambda_1 - \mu + \nu; 2\,(1 - \mu); -e^{-a L}\right)$ \\
$h_5$    & ${}_2F_1\left(\dfrac{1}{2}-\lambda - \mu - \nu, \dfrac{1}{2}+\lambda - \mu - \nu; 1 - 2\nu; \dfrac{1}{1+e^{-a L}}\right)$ \\
$h_6$    & ${}_2F_1\left(\dfrac{3}{2}-\lambda - \mu - \nu, \dfrac{3}{2}+\lambda - \mu - \nu; 2\,(1 - \nu); \dfrac{1}{1+e^{-a L}}\right)$ \\ 
$h_7$    & ${}_2\tilde{F}_1\left(\dfrac{3}{2}-\lambda - \mu - \nu, \dfrac{1}{2}+\lambda - \mu - \nu; 1 - 2\nu; \dfrac{1}{1+e^{-a L}}\right)$\\
$h_8$    & ${}_2\tilde{F}_1\left(\lambda_1 - \mu - \nu,-\lambda_1 - \mu + \nu; 1 - 2\mu; -e^{-a L}\right)$\\
$h_9$    & ${}_2\tilde{F}_1\left(\dfrac{1}{2}-\lambda - \mu - \nu, \dfrac{1}{2}+\lambda - \mu - \nu; 1 - 2\nu; \dfrac{1}{1+e^{-a L}}\right)$ \\
$h_{10}$ & ${}_2\tilde{F}_1\left(-\lambda_1 - \mu - \nu, -\lambda_1 - \mu + \nu; 1 - 2\mu; -e^{-a L}\right)$\\
$h_{11}$ & ${}_2\tilde{F}_1\left(1-\lambda_1 - \mu - \nu, -\lambda_1 - \mu + \nu; 1 - 2\mu;-e^{-a L}\right)$\\
$h_{12}$    & ${}_2{F}_1\left(1-\lambda_1 -\mu -\nu,-\lambda_1 -\mu +\nu;1-2 \mu;-e^{-a L}\right)$\\
$h_{13}$   & ${}_2{F}_1\left(1-\lambda_1+\mu -\nu,-\lambda_1+\mu +\nu;1+2 \mu;-e^{-a L}\right)$\\
\bottomrule
\end{tabular}
\label{tab:hypergeometric_functions}
\end{table}

Here, from the system of equations  Eq. \eqref{eq052}, 
Eq. \eqref{eq053}, and  Eq. \eqref{eq3}  the coefficients $D_1$, $D_2$ are obtained in terms of $d_1$. Using some forms and properties of the Gamma functions $\Gamma(x)$ and the functions $h_i$ in Table \ref{tab:hypergeometric_functions}, they can be read as,

\begin{align}
D_1 &=\dfrac{
\begin{aligned}
& + d_1 \,\Bigg\{\Gamma (1-2 \mu )\, \Gamma (1-2 \nu ) \left(-e^{-a L}\right)^{-\mu } 
\left(1+e^{-a L}\right)^{\nu+\lambda_1} \left(\dfrac{1}{1+e^{a L}}\right)^{1-\mu} \\
& \quad \times \Bigg[ -\left( -1+2\lambda+2\mu+2\nu\right) h_7 h_8 \\
& \quad + \left[ -1+2\lambda+2\mu-2\nu-2 e^{aL} \left(\lambda_1-\mu+\nu \right)\right]  h_9 h_{10} \\
& \quad + 2 \left(1+e^{aL}\right) \left( \lambda_1+\mu+\nu\right) h_{9} h_{11} \Bigg] \Bigg\}
\end{aligned}
}{
\begin{aligned}
& 2 \,\Bigg[\left(\lambda_1+\mu+\nu \right) h_1 h_{12} - \left(\lambda_1-\mu+\nu\right) h_{4} h_{13} \Bigg]
\end{aligned}
}.
\end{align}

\vspace{5cm}
The same shape can be made approximately for $D_2$,

\begin{equation}
D_2 = \dfrac{
\begin{aligned}
&+ d_1 \left(-e^{-a L}\right)^{\mu} \left(1+e^{-a L}\right)^{\nu+\lambda_1}
\left(\dfrac{1}{1+e^{a L}}\right)^{1-\mu} \\
& \quad \times \Bigg\{ \left(-1+2 \lambda+2 \mu+ 2\nu\right) h_{1} h_{7} \\
& \quad - \left[1 - 2\lambda - 2\mu + 2\nu + 2 e^{aL} \left(\lambda_1-\mu+\nu \right) \right] h_{1} h_{5}  \\
& \quad - 2\left(1+e^{aL}\right)\left(\lambda_1-\mu+\nu\right) h_{1} h_{13}  \Bigg\}
\end{aligned}
}{
\begin{aligned}
& 2\, \Bigg[\left(\lambda_1+\mu+\nu \right) h_{1} h_{11} - \left(\lambda_1-\mu+\nu\right) h_{3} h_{13} \Bigg]
\end{aligned}
}.
\label{eq056}
\end{equation}

\begin{figure}[th!]
\centering
\includegraphics[scale=0.7]{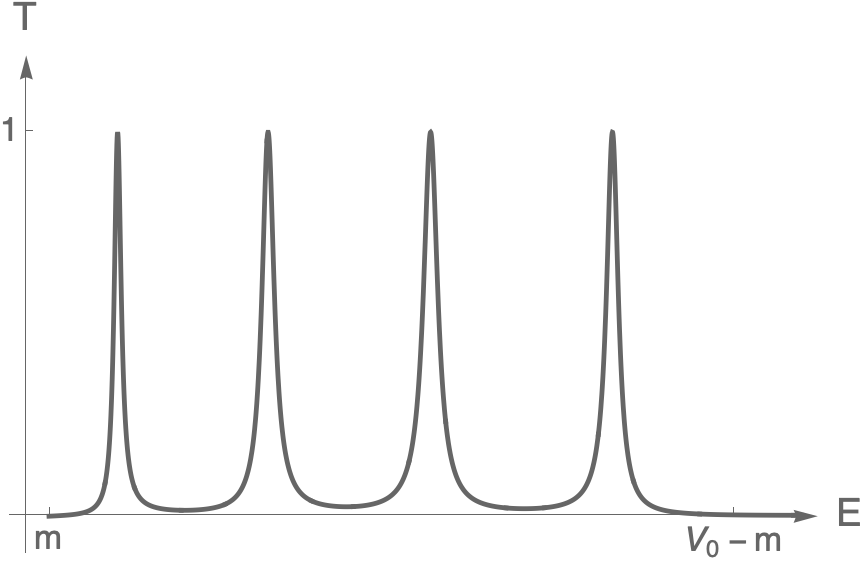}
\caption{Transmission coefficient $T$ for the Woods--Saxon potential barrier considering  $a = 2$, $L = 2$, and $V_0 = 5$ in function of the energy $E$.  The four transmission resonances $T=1$ are located at at: a) $E=1.30086$, b) $E=1.96049$, c) $E=2.67321$, and d)  $E=3.47053$.}
\label{T_1}
\end{figure}

\begin{figure}[th!]
\centering
\includegraphics[scale=0.7]{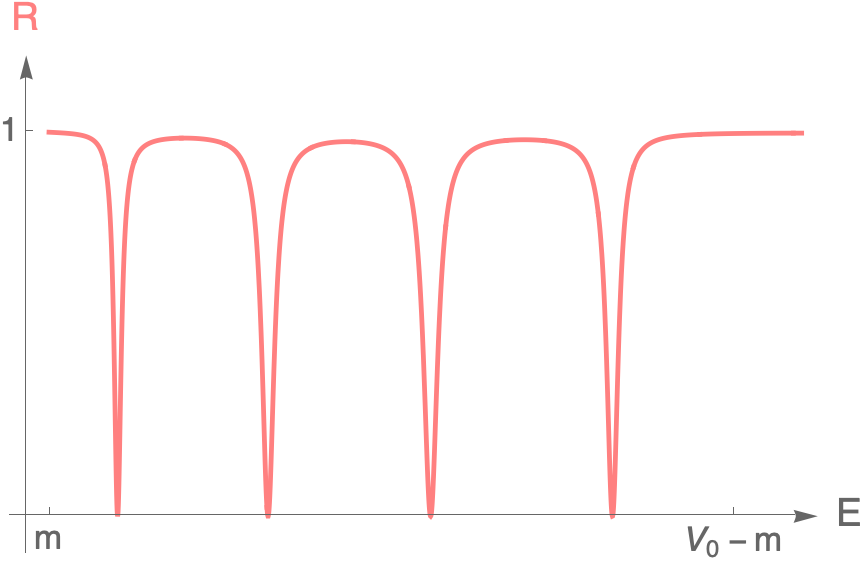}
\caption{Reflection coefficient $R$ for the Woods--Saxon potential barrier considering $a = 2$, $L = 2$, and $V_0 = 5$ in function of the energy $E$.  The four values of the energy $E$ where $R=0$ are located at at: a) $E=1.30086$, b) $E=1.96049$, c) $E=2.67321$, and d)  $E=3.47053$.}
\label{R_1}
\end{figure}

\begin{figure}[th!]
\centering
\includegraphics[scale=0.65]{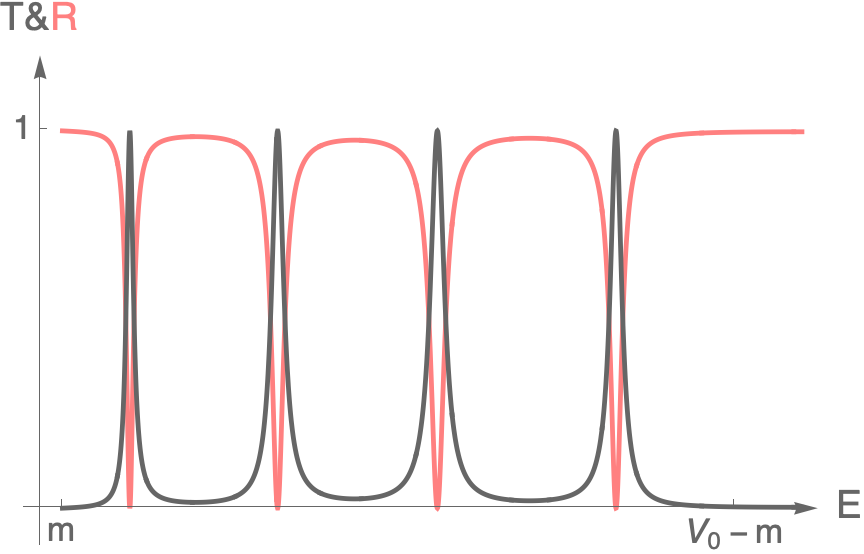}
\caption{Transmission $T$ and Reflection $R$ coefficients for the Woods--Saxon potential barrier considering $a = 2$, $L = 2$, and $V_0 = 5$ in function of the energy $E$. Solid gray line represents the transmission coefficient $T$, and solid pink line represents the reflection coefficient $R$. The four transmission resonances $T=1$ are located at at: a) $E=1.30086$, b) $E=1.96049$, c) $E=2.67321$, and d)  $E=3.47053$, and when $T=1$ then $R=0$ because the unitary condition.}
\label{TR_1}
\end{figure}

\bigskip\bigskip
Using the new constants $D_1$, and $D_2$ inside the definitions of $A$ and $B$ in terms of Gamma functions $\Gamma(x)$, such as Eq. (\ref{eq041}), we can calculate the transmission coefficient $T$ and the reflection coefficient $R$ using Eqs. \eqref{eq045b} and \eqref{eq045a}, respectively. In Fig. \ref{T_1} we can observed  the transmission coefficient $T$ for $a = 2$, $L = 2$, and $V_0 = 5$ as a function of the energy $E$. Also, we can observed that we have four transmission resonances $T=1$ at: a) $E=1.30086$, b) $E=1.96049$, c) $E=2.67321$, and d)  $E=3.47053$. Fig. \ref{R_1} shows the reflection coefficient $R$ for the same case, we can see that where $T=1$ then $R=0$, and the unitary condition $T+R=1$ is satisfied. In Fig. \ref{TR_1} we can observe both cases together.

\begin{figure}[th!]
\centering
\includegraphics[scale=0.7]{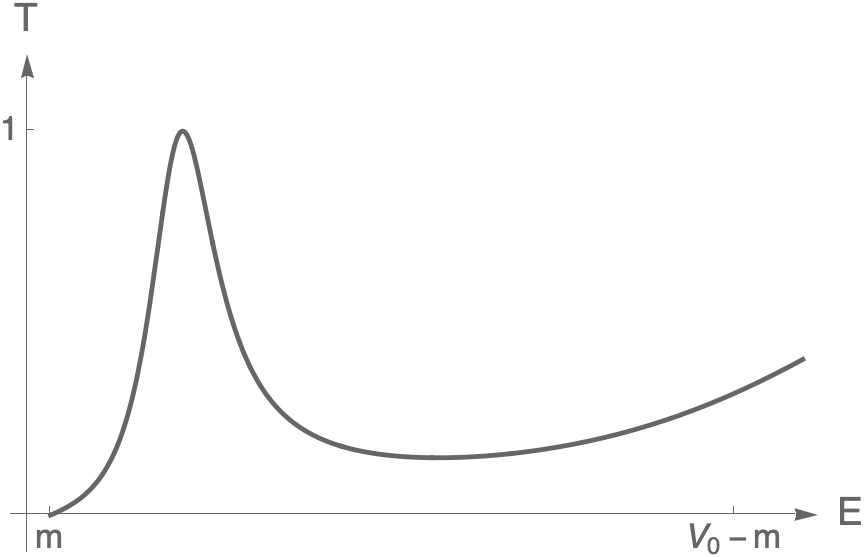}
\caption{Transmission coefficient $T$ for the Woods--Saxon potential barrier considering  $a = 3$, $L = 0.4$, and $V_0 = 5$ in function of the energy $E$. For this set of parameters we have only one transmission resonance $T = 1$ at $E = 1.57991$.}
\label{T_2}
\end{figure}

\begin{figure}[th!]
\centering
\includegraphics[scale=0.7]{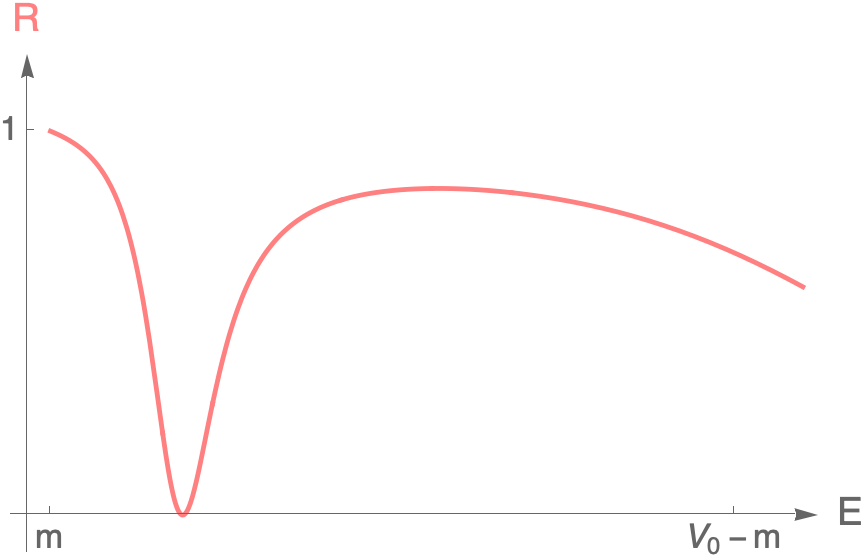}
\caption{Reflection coefficient $R$ for the Woods--Saxon potential barrier considering $a = 3$, $L = 0.4$, and $V_0 = 5$ in function of the energy $E$. For this set of parameters $R = 0$ at $E = 1.57991$.}
\label{R_2}
\end{figure}

\begin{figure}[th!]
\centering
\includegraphics[scale=0.7]{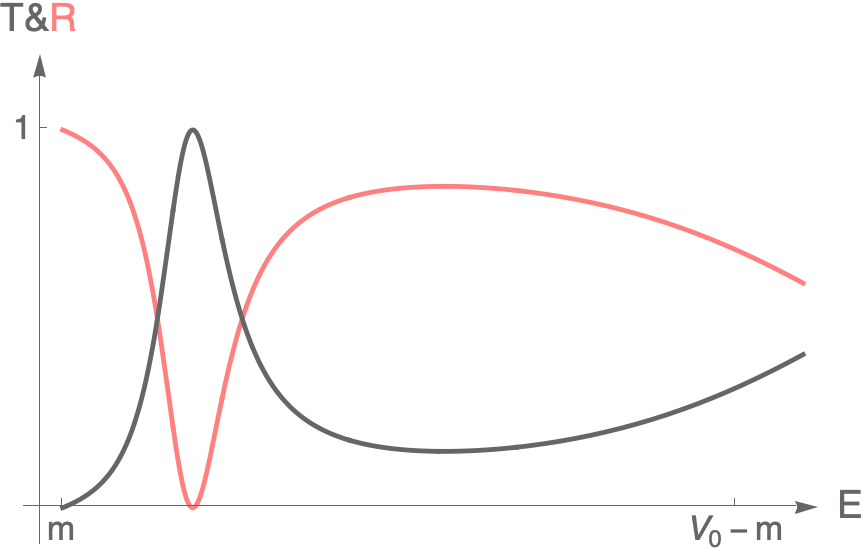}
\caption{Transmission $T$ and Reflection $R$ coefficients for the Woods--Saxon potential barrier considering $a = 3$, $L = 0.4$, and $V_0 = 5$ in function of the energy $E$. Solid gray line represents the transmission coefficient $T$, and solid pink line represents the reflection coefficient $R$. We can observed that there is  only one transmission resonance $T = 1$ at $E = 1.57991$, and $R=0$ at the same value of energy.}
\label{TR_2}
\end{figure}

Fig. \ref{T_2} shows the transmission coefficient $T$ for $a = 3$, $L = 0.4$, and $V_0 = 5$ as a function of the energy $E$. We can observed that we have only one transmission resonance $T=1$ at $E=1.57991$. Fig. \ref{R_2} shows the reflection coefficient $R$ for the same case, we can see again that the unitary condition $T+R=1$ is satisfied. In Fig. \ref{TR_2} we can observe both cases together.

\subsection{Transmission resonances in the square potential barrier vs. the Woods--Saxon potential barrier}

\begin{figure}[htbp]
\centering
\includegraphics[scale=0.6]{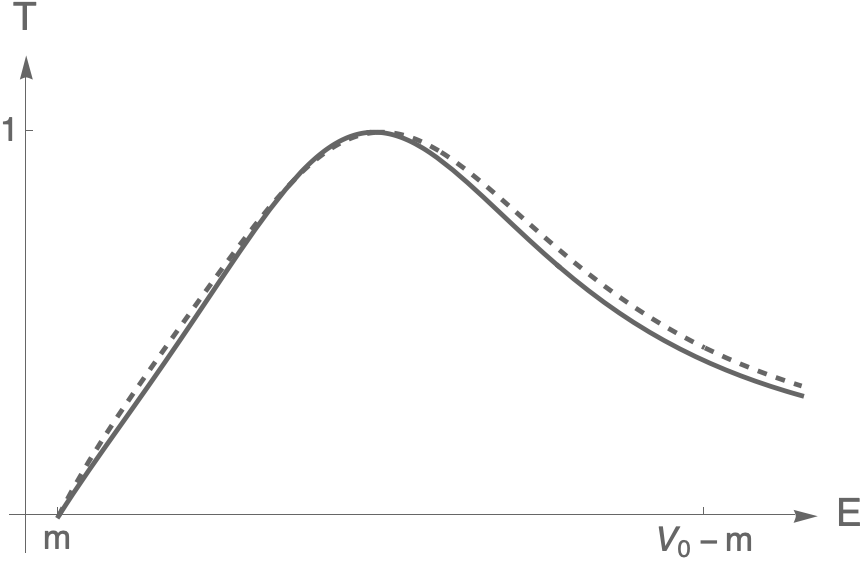}
\caption{Transmission coefficient $T$ for the Woods--Saxon potential barrier with $a = 70$, $L = 0.4$, and $V_0 = 4$, and for the square potential barrier with $L = 0.4$, and $V_0 = 4$ in function of the energy $E$. Solid gray line represents the Woods--Saxon potential barrier, and dashed gray line represents the square potential barrier. In the figure you can observed that we only have a transsmission resonance $T=1$ at $E=1.97899$ for the Woods--Saxon potential barrier, and at $E=2.00000$ for the square potential barrier}.
\label{T_3}
\end{figure}

\begin{figure}[htbp]
\centering
\includegraphics[scale=0.6]{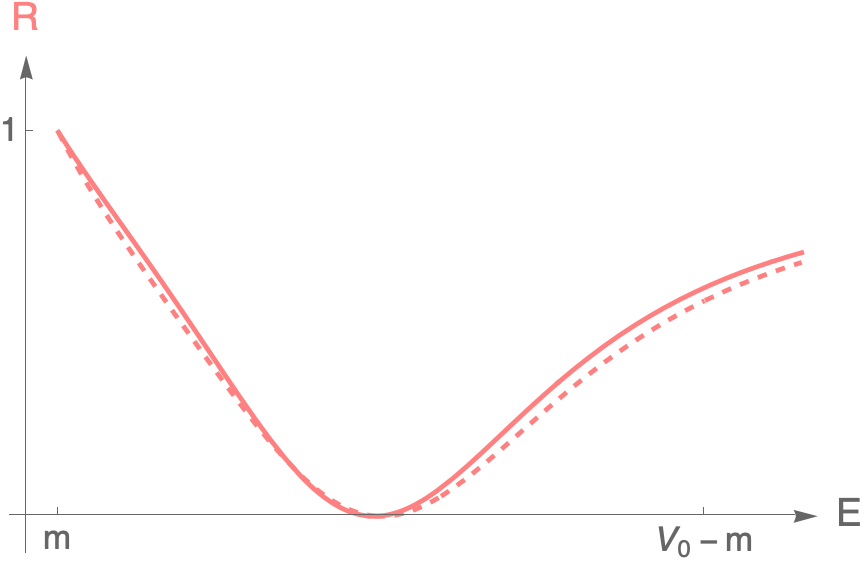}
\caption{Reflection coefficient $R$ for the Woods--Saxon potential barrier with $a = 70$, $L = 0.4$, and $V_0 = 4$, and for the square potential barrier with $L = 0.4$, and $V_0 = 4$ in function of the energy $E$. Solid pink line represents the Woods--Saxon potential barrier, and dashed pink line represents the square potential barrier. In the figure you can observed that $R=0$ at $E=1.97899$ for the Woods--Saxon potential barrier, and at $E=2.00000$ for the square potential barrier}.
\label{R_3}
\end{figure}

In this subsection, we show the plots for the transmission coefficient $T$ and the reflection coefficient $R$, but using a higher value for $a=70$, a lower value for $L=0.4$, and $V_0=4$ in the Woods--Saxon potential barrier in order to compare with those obtained by the square potential barrier. In Figs. \ref{T_3} and \ref{R_3} we can observe the transmission coefficient $T$ and the reflection coefficient $R$ for this case compared our results for one of the square potential barrier. 

The position of the transmission resonance in each case is shown in Table \ref{table:scattering}, from which  we can say that the position of the transmission resonance is slightly move for the Wood--Saxon potential barrier compared to the square potential barrier. In order to obtain the same transmission $T$ and reflection $R$ coefficients we need to make the Woods--Saxon potential barrier tends to the square potential barrier by moving higher the value of $a$ together with the value of $L$.

\begin{table}[thpb]
\begin{center}
\caption{Comparison of the position of the transmission resonance for the Woods--Saxon potential barrier with $a = 70$, $L = 0.4$, and $V_0 = 4$, and the square potential barrier with $L = 0.4$, and $V_0 = 4$.}
\begin{tabular}{c|c|c}
\toprule
Potential barrier & Transmission coefficient $T$ & Energy $E$  \\
\midrule 
Woods--Saxon  & 1 & 1.97899   \\ 
\midrule
Square  & 1 & 2.00000    \\ 
\bottomrule
\end{tabular}
\label{table:scattering}
\end{center}
\end{table}

\subsection{Transmission resonances in the Cups potential barrier vs. the Woods--Saxon potential barrier}

\begin{figure}[th!]
\centering
\includegraphics[scale=0.58]{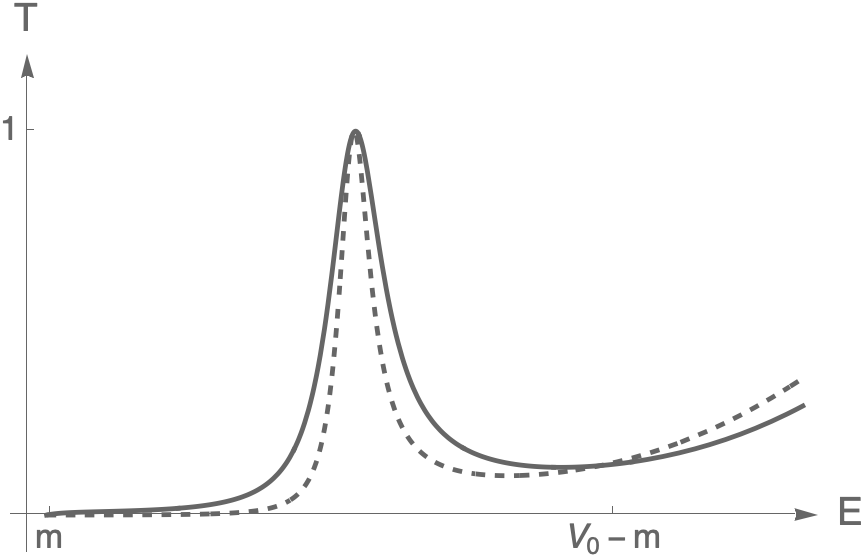}
\caption{Transmission coefficient $T$ for the Woods--Saxon potential barrier with $a=2$, $L=0.4$, and $V_0=7.24$, and for the cusp potential barrier with $a=1.2$, and $V_0=5$ in function of the energy $E$. Solid gray line represents the Woods--Saxon potential barrier, and dashed gray line represents the cusp potential barrier. In the figure you can observed that we only have a transsmission resonance $T=1$ at $2.63455$ for the Woods--Saxon potential barrier, and at $E=2.62679$ for the cusp potential barrier.}
\label{T_4}
\end{figure}

\begin{figure}[th!]
\centering
\includegraphics[scale=0.58]{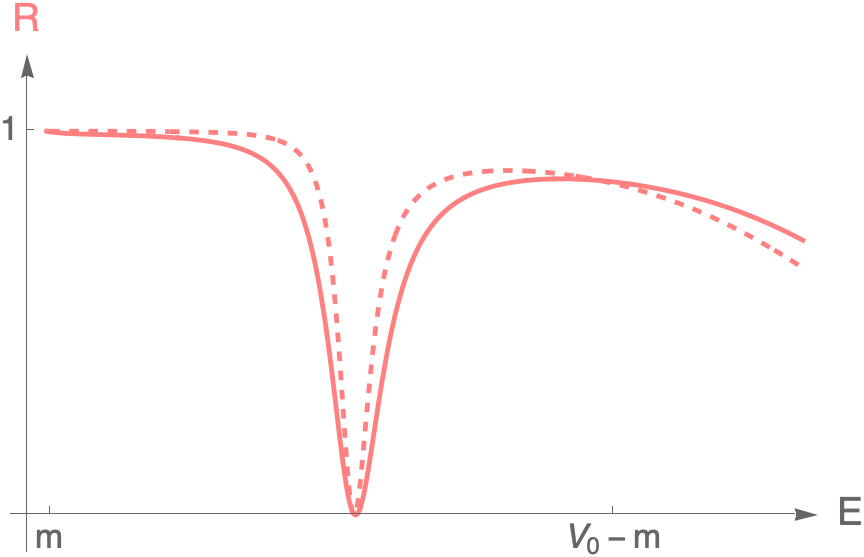}
\caption{Reflection coefficient $R$ for the Woods--Saxon potential barrier with $a=2$, $L=0.4$, and $V_0=7.24$, and for the cusp potential barrier with $a=1.2$, and $V_0=5$ in function of the energy $E$. Solid gray line represents the Woods--Saxon potential barrier, and dashed gray line represents the cusp potential barrier. In the figure you can observed that $R=0$ at $E=2.63455$ for the Woods--Saxon potential barrier, and at $E=2.62679$  for the cusp potential barrier.}
\label{R_4}
\end{figure}

In this subsection, we present the plots of the transmission coefficient $T$ and the reflection coefficient $R$, using the parameters $a=2$, $L=0.4$, and $V_0=7.24$, in order to compare our results with those obtained for the cusp potential barrier  \cite{rojas:2024}.  The parameters for the Woods--saxon potential barrier are chosen in this way because it closely approximates the shape of a cusp potential barrier with parameters $a=1.2$, and $V_0=5$. Figures \ref{T_4} and \ref{R_4} show the transmission $T$ and reflection $R$ coefficients  for this case, allowing for a direct comparison between both potential barriers.

\newpage
The positions of the transmission resonances for each case are showed in Table \ref{table:scattering_cuspWS}, from which we observe that the resonance transmission position is slightly shifted for the Woods-?Saxon potential barrier compared to the cusp potential barrier. To obtain the same transmission $T$ and reflection $R$ coefficients, the Woods--Saxon potential must approach better the cusp potential barrier. This can be achieved by adjusting the values of $a$, $L$, and $V_0$ for the Woods--Saxon potential barrier.

\begin{table}[thpb]
\begin{center}
\caption{Comparison of the position of the transmission resonance for the Woods--Saxon potential barrier with $a=2$, $L=0.4$, and $V_0=7.24$, and the cusp potential barrier with $a=1.2$, and $V_0=5$.}
\begin{tabular}{c|c|c}
\toprule
Potential barrier & Transmission coefficient $T$ & Energy $E$  \\
\midrule 
Woods--Saxon  & 1 & 2.63455    \\ 
\midrule
Cusp  & 1 & 2.62679   \\ 
\bottomrule
\end{tabular}
\label{table:scattering_cuspWS}
\end{center}
\end{table}

\section{Bound States}

To solve the problem, it is important to take into account the relation between the positive ($R$) and negative ($L$) regions of the Klein--Gordon equation,

\begin{equation}
\label{eq14}
\dfrac{\D^2\phi_1{_{LR}}(x)}{\D x^2} + \left\{[E - V(x)]^2 - 1 \right\} \phi_1{_{LR}}(x) = 0.
\end{equation}

In addition, the potential is divided into two regions as shown in Eq. \eqref{eq002_well}.

First of all, we are going to work with the first component of the spinor.

\subsection{For $x<0$}

\subsubsection{Solution for the first component of the spinor (left region)}

\begin{equation}
\label{bs_eq045}
\dfrac{\D^2\phi_{1L}(x)}{\D x^2} + \left\{\left[E + \dfrac{V_0}{1+e^{-a(x+L)}}\right]^2 - 1 \right\} \phi_{1L}(x) = 0,
\end{equation}
applying the change of variable $y^{-1}=1+e^{-a(x+L)}$ into Eq. (\ref{bs_eq045}) we obtain \cite{rojas2006klein},

\begin{equation}
\label{bs_eq046}
a^2y(1-y)\dfrac{\D}{\D y}\left[y(1-y)\dfrac{\D \phi_{1L}(y)}{\D y}\right] + \left[\left(E + V_0 y\right)^2 - 1 \right] \phi_{1L}(y) = 0,
\end{equation}
then, defining the function $\phi_{1L}(y) = y^{\sigma}(1-y)^{\gamma}h(y)$, we get

\begin{equation}
\label{bs_eq047}
y(1-y) \, h'' + [(1+2\sigma) - 2(\sigma+\gamma+1)y] \,h' -\left(\dfrac{1}{2} + \sigma + \gamma + \lambda\right)\left(\dfrac{1}{2} + \sigma + \gamma - \lambda\right) \, h = 0,
\end{equation}
where prime indicates derivative respect to $y$ and,

\begin{eqnarray}
\label{eq048}
\sigma &=& \dfrac{\sqrt{1-E^2}}{a},\\
\gamma &=& \dfrac{\sqrt{1-(E+V_0)^2}}{a},\\
\lambda &=& \dfrac{\sqrt{a^2 - 4V_0^2}}{2a}.
\end{eqnarray}

Eq. (\ref{bs_eq047}) has the general solution,

\begin{eqnarray}
\label{bs_eq049}
\nonumber
h(y) &=& a_{1} \, {}_2F_1\left(\dfrac{1}{2} + \gamma-\lambda+\sigma, \dfrac{1}{2} + \gamma+\lambda+\sigma; 1+2\sigma; y\right) \\
&+& a_2 \, y^{-2\sigma} \ {}_2F_1\left(\dfrac{1}{2} + \gamma-\lambda-\sigma, \dfrac{1}{2} + \gamma+\lambda-\sigma; 1-2\sigma; y\right),
\end{eqnarray}
then, we can write the expression for $\phi_{1L}(y)$,

\begin{eqnarray}
\nonumber
\phi_{1L}(y) &=& a_{1}\, y^\sigma (1-y)^{\gamma}\ {}_2F_1\left(\dfrac{1}{2} + \gamma-\lambda+\sigma, \dfrac{1}{2} + \gamma+\lambda+\sigma; 1+2\sigma; y\right) \\
&+& a_2 \,y^{-\sigma}(1-y)^{\gamma} \ {}_2F_1\left(\dfrac{1}{2} + \gamma-\lambda-\sigma, \dfrac{1}{2} + \gamma+\lambda-\sigma; 1-2\sigma; y\right),
\label{bs_eq050}
\end{eqnarray}
where $a_{1}$ and $a_{2}$ are arbitrary constants.

\subsection{For $x>0$}

\subsubsection{Solution for the first component of the spinor (right region)}

\begin{equation}
\label{bs_eq051}
\dfrac{\D^2\phi_{1R}(x)}{\D x^2} + \left\{\left[E + \dfrac{V_0}{1+e^{a(x-L)}}\right]^2 - 1 \right\} \phi_{1R}(x) = 0,
\end{equation}
applying the substitution $z^{-1}=1+e^{a(x-L)}$ into Eq. (\ref{bs_eq051}) as in \cite{rojas2006klein}, we obtain,

\begin{equation}
\label{bs_eq052}
(1-z)z \ a^2 \ \dfrac{\D}{\D z} \left[(1-z)z \ \dfrac{\D \phi_{1R}(z)}{\D z}\right] + \left[\left(E + V_0 \ z\right)^2 - 1 \right] \phi_{1R}(z) = 0,
\end{equation}
then, we substitute $\phi_{1R}(z) = z^{\sigma}(1-z)^{-\gamma}g(z)$ into Eq. (\ref{bs_eq052}),

\begin{equation}
\label{bs_eq053}
z(1-z) \ g'' + [(1+2\sigma) - 2(\sigma-\gamma+1)z] \ g' -\left(\dfrac{1}{2} - \gamma + \lambda + \sigma\right)\left(\dfrac{1}{2}  - \gamma - \lambda + \sigma\right) \ g = 0,
\end{equation}
where the primes denote derivatives respect to $z$.
Eq. (\ref{bs_eq053}) has the general solution,

\begin{eqnarray}
\label{bs_eq055}
\nonumber
g(z) &=& b_{1} \, {}_2F_1\left(\dfrac{1}{2} - \gamma-\lambda+\sigma, \dfrac{1}{2} - \gamma+\lambda+\sigma; 1+2\sigma; z\right) \\
&+&  b_2 \, z^{-2\sigma} \ {}_2F_1\left(\dfrac{1}{2} - \gamma-\lambda-\sigma, \dfrac{1}{2} - \gamma+\lambda-\sigma; 1-2\sigma; z\right),
\end{eqnarray}
then, we can write the expression for $\phi_{1R}(z)$,

\begin{eqnarray}
\label{bs_eq056}
\phi_{1R}(z) &=& b_{1} \, z^\sigma (1-z)^{-\gamma}\ {}_2F_1\left(\dfrac{1}{2} - \gamma-\lambda +\sigma, \dfrac{1}{2} - \gamma+\lambda+\sigma; 1+2\sigma; z\right) \nonumber\\
&+&b_2 \, z^{-\sigma}(1-z)^{-\gamma} \ {}_2F_1\left(\dfrac{1}{2} - \gamma-\lambda-\sigma, \dfrac{1}{2} - \gamma+\lambda-\sigma; 1-2\sigma; z\right),
\end{eqnarray}
where $b_{1}$ and $b_{2}$ are arbitrary constants.

\subsection{Asymptotic behavior}

We study the asymptotic behavior for Eqs. (\ref{bs_eq050}) and (\ref{bs_eq056}), as $x\rightarrow-\infty$, $y \rightarrow 0$ and as $x\rightarrow\infty$, $z\rightarrow0$. As a result, we obtain the following regular wave functions,

\begin{eqnarray}
\label{bs_eq057}
\phi_{1L}(y) &=& a_{1}\, y^\sigma (1-y)^{\gamma}\ {}_2F_1\left(\dfrac{1}{2} + \gamma-\lambda+\sigma, \dfrac{1}{2} + \gamma+\lambda+\sigma; 1+2\sigma; y\right),\\
\label{bseq058}
\phi_{1R}(z) &=& b_{1}\, z^\sigma (1-z)^{-\gamma}\ {}_2F_1\left(\dfrac{1}{2} - \gamma-\lambda+\sigma, \dfrac{1}{2} - \gamma+\lambda+\sigma; 1+2\sigma; z\right).
\end{eqnarray}

Now in terms of $x$,

\begin{eqnarray}
\nonumber
\label{bs_eq059}
\phi_{1L}(x) &=& a_{1} \left[\dfrac{1}{1+e^{-a(x+L)}}\right]^\sigma \left[1-\dfrac{1}{1+e^{-a(x+L)}}\right]^{\gamma}\\
&\times& {}_2F_1\left[\dfrac{1}{2} \gamma-\lambda+\sigma, \dfrac{1}{2} + \gamma+\lambda+\sigma; 1+2\sigma; \dfrac{1}{1+e^{-a(x+L)}}\right],\\
\nonumber
\label{bs_eq060}
\phi_{1R}(x) &=& b_{1} \left[\dfrac{1}{1+e^{a(x-L)}}\right]^\sigma \left[1-\dfrac{1}{1+e^{a(x-L)}}\right]^{-\gamma}\\
&\times& {}_2F_1\left[\dfrac{1}{2} - \gamma-\lambda+\sigma, \dfrac{1}{2} - \gamma+\lambda+\sigma; 1+2\sigma; \dfrac{1}{1+e^{a(x-L)}}\right].
\end{eqnarray}

\subsection{Solution for the second and third components of the spinor}

Once we obtain the solutions related to the Klein--Gordon equation, we can calculate the wave spinor components of the DKP equation. These components are expressed in equation (\ref{Dkpequation_system}).

Based on equation (\ref{eq007}), we get the following results for $x < 0$, as in reference \cite{rojas:2024},

\begin{equation}\label{eq32}
\phi_L(x)= \begin{pmatrix}
\phi_{1L}(x) \\
\phi_{2L}(x) \\
\phi_{3L}(x) \\
\end{pmatrix} ,
\end{equation}
where the  components of the spinor are the following,

\begin{eqnarray}
\nonumber
\phi_{1L}(x) &=& a_{1} \left[\dfrac{1}{1+e^{-a(x+L)}}\right]^\sigma \left[1-\dfrac{1}{1+e^{-a(x+L)}}\right]^{\gamma}\\
&\times& {}_2F_1\left[\dfrac{1}{2} + \gamma-\lambda+\sigma, \dfrac{1}{2} + \gamma+\lambda+\sigma; 1+2\sigma; \dfrac{1}{1+e^{-a(x+L)}}\right], \label{eq33}\\
\nonumber
\phi_{2L}(x) &=& a_{1}\, a\, e^{-a(x+L)}  \left[\dfrac{1}{1+e^{-a(x+L)}}\right]^{2 + \sigma} \left[\dfrac{1}{1+e^{a(x+L)}}\right]^{\gamma} \\
\nonumber
&\times& \Bigg\{\left[\gamma + e^{a(x+L)}\gamma - \sigma -e^{-a(x+L)}\sigma\right]   \nonumber \\
& \times&  {}_2F_1\left[\dfrac{1}{2} + \gamma-\lambda+\sigma, \dfrac{1}{2} + \gamma+\lambda+\sigma; 1+2\sigma; \dfrac{1}{1+e^{-a(x+L)}}\right]  \nonumber \\
\nonumber
& -& \dfrac{1}{4}\left(\dfrac{1}{1 + 2\sigma}\right) \Bigg[(1+2\gamma - 2\lambda+2\sigma)(1+2\gamma+2\lambda+2\sigma)  \\
\nonumber
&\times&   {}_2F_1\left[\dfrac{3}{2}+\gamma-\lambda+\sigma,\dfrac{3}{2}+\gamma +\lambda+\sigma;2\,(1+\sigma);\dfrac{1}{1+e^{-a(x+L)}}\right]\Bigg]\Bigg\},
\label{eq34}\\\\
\nonumber
\phi_{3L}(x) &=& - i \,a_{1} \,\left[\dfrac{1}{1+e^{-a(x+L)}}\right]^\sigma \left[1-\dfrac{1}{1+e^{-a(x+L)}}\right]^{\gamma} \left[E+\dfrac{V_0}{1+e^{-a(x+L)}}\right]\\ 
& \times&    {}_2F_1\left[\dfrac{1}{2} + \gamma-\lambda+\sigma, \dfrac{1}{2} + \gamma+\lambda+\sigma; 1+2\sigma; \dfrac{1}{1+e^{-a(x+L)}}\right]. \label{eq35}
\end{eqnarray}

Following Eq. (\ref{eq007}) for $x > 0$ we have,

\begin{equation}
\label{eq36}
\phi_{R}(x)= \begin{pmatrix}
\phi_{1R}(x) \\
\phi_{2R}(x) \\
\phi_{3R}(x) \\
\end{pmatrix} ,
\end{equation}

\noindent where the components of the spinor are expressed as follows,

\begin{eqnarray}
\nonumber
\phi_{1R}(x) &=& b_{1}\, \left[\dfrac{1}{1+e^{a(x-L)}}\right]^\sigma \left[1-\dfrac{1}{1+e^{a(x-L)}}\right]^{-\gamma}\\
&\times& {}_2F_1\left[\dfrac{1}{2} - \gamma-\lambda+\sigma, \dfrac{1}{2} - \gamma+\lambda+\sigma; 1+2\sigma; \dfrac{1}{1+e^{a(x-L)}}\right], \\
\nonumber
\phi_{2R}(x) &=& b_{1}\,  a \, e^{a(x-L)}  \left[\dfrac{1}{1+e^{a(x-L)}}\right]^{2 + \sigma} \left[\dfrac{1}{1+e^{a(L-x)}}\right]^{-\gamma}\\
\nonumber
&\times& \Bigg\{\left[\gamma + e^{a(-x+L)}\gamma + \sigma +e^{a(x-L)}\sigma\right] \nonumber\\
\nonumber
& \times&  {}_2F_1\left[\dfrac{1}{2} - \gamma-\lambda+\sigma, \dfrac{1}{2} - \gamma+\lambda+\sigma; 1+2\sigma; \dfrac{1}{1+e^{a(x-L)}}\right]\\ 
\nonumber
& +& \dfrac{1}{4} \left(\dfrac{1}{1 + 2\sigma}\right) \Bigg[(-1+2\gamma - 2\lambda-2\sigma)(-1+2\gamma+2\lambda-2\sigma)  \\
\nonumber
&\times& 
{}_2F_1\left[\dfrac{3}{2}-\gamma-\lambda+\sigma,\dfrac{3}{2}-\gamma +\lambda+\sigma;2\, (1+\sigma);\dfrac{1}{1+e^{a(x-L)}}\right]\Bigg]\Bigg\}, 
\nonumber
\\\\
\nonumber 
\phi_{3R}(x) &=& - i \, b_{1} \,\left[\dfrac{1}{1+e^{a(x-L)}}\right]^\sigma\left[1-\dfrac{1}{1+e^{a(x-L)}}\right]^{-\gamma}\ \left[E+\dfrac{V_0}{1+e^{a(x-L)}}\right] \\
& \times&  {}_2F_1\left[\dfrac{1}{2} - \gamma-\lambda+\sigma, \dfrac{1}{2} - \gamma+\lambda+\sigma; 1+2\sigma; \dfrac{1}{1+e^{a(x-L)}}\right]. \label{eq39}
\end{eqnarray}

It is possible to find the bound states using the condition $|E|<1$ and analyzing the continuity at $x=0$ of the spinor for both regions \cite{sogut:2010},

\begin{equation}
\phi_L (x) \Big|_{x = 0}  = \phi_R (x) \Big|_{x = 0}.
\label{eq40}
\end{equation}

The condition of equation (\ref{eq40}) allows us to get for the left side,

\begin{eqnarray}
\nonumber
\phi_{1L}(0) &=& a_{1} \left(\dfrac{1}{1+e^{-aL}}\right)^\sigma \left(1-\dfrac{1}{1+e^{-aL}}\right)^{\gamma}\ \\
&\times&
{}_2F_1\left(\dfrac{1}{2} + \gamma-\lambda+\sigma, \dfrac{1}{2} + \gamma+\lambda+\sigma; 1+2\sigma; \dfrac{1}{1+e^{-aL}}\right), 
\end{eqnarray}

\begin{eqnarray}
\nonumber
\phi_{2L}(0) &=& a\ a_{1}\, e^{-aL}  \left(\dfrac{1}{1+e^{-aL}}\right)^{2 + \sigma} \left(\dfrac{1}{1+e^{aL}}\right)^{\gamma}\ \Bigg\{\left(\gamma + e^{aL}\gamma - \sigma -e^{-aL}\sigma\right)  \nonumber \\
\nonumber
& \times& {}_2F_1\left(\dfrac{1}{2} + \gamma-\lambda+\sigma, \dfrac{1}{2} + \gamma+\lambda+\sigma; 1+2\sigma; \dfrac{1}{1+e^{-aL}}\right) \nonumber \\ 
\nonumber
& -&   \dfrac{1}{4}\left(\dfrac{1}{1 + 2\sigma}\right)\Bigg[(1+2\gamma - 2\lambda+2\sigma)(1+2\gamma+2\lambda+2\sigma) 
\\
\nonumber
&\times&  
{}_2F_1\left(\dfrac{3}{2}+\gamma-\lambda+\sigma,\dfrac{3}{2}+\gamma +\lambda+\sigma;2\,(1+\sigma);\dfrac{1}{1+e^{-aL}}\right)\Bigg]\Bigg\},
\nonumber \\\\
\phi_{3L}(0) &=& - i\, a_{1} \,\left(\dfrac{1}{1+e^{-aL}}\right)^\sigma \left(1-\dfrac{1}{1+e^{-aL}}\right)^{\gamma} \left(E+\dfrac{V_0}{1+e^{-aL}}\right)\nonumber\\ 
& \times&  {}_2F_1\left(\dfrac{1}{2} + \gamma-\lambda+\sigma, \dfrac{1}{2} + \gamma+\lambda+\sigma; 1+2\sigma; \dfrac{1}{1+e^{-aL}}\right). 
\nonumber\\
\end{eqnarray}

\bigskip
And for the right side, it is,

\bigskip
\begin{eqnarray}
\nonumber
\phi_{1R}(0) &=& b_{1} \,\left(\dfrac{1}{1+e^{-aL}}\right)^\sigma \left(1-\dfrac{1}{1+e^{-aL}}\right)^{-\gamma}\\
&\times& {}_2F_1\left(\dfrac{1}{2} - \gamma-\lambda+\sigma, \dfrac{1}{2} - \gamma+\lambda+\sigma; 1+2\sigma; \dfrac{1}{1+e^{-aL}}\right), \\
\nonumber
\phi_{2R}(0) &=& b_{1}\, a \,e^{-aL}  \left(\dfrac{1}{1+e^{-aL}}\right)^{2 + \sigma} \left(\dfrac{1}{1+e^{aL}}\right)^{-\gamma}\ \Bigg\{\left(\gamma + e^{aL}\gamma + \sigma +e^{-aL}\sigma\right) \\
\nonumber
& \times& {}_2F_1\left(\dfrac{1}{2} - \gamma-\lambda+\sigma, \dfrac{1}{2} - \gamma+\lambda+\sigma; 1+2\sigma; \dfrac{1}{1+e^{-aL}}\right) \nonumber \\ 
\nonumber
& +& \dfrac{1}{4}\left(\dfrac{1}{1 + 2\sigma}\right)\Bigg[(-1+2\gamma - 2\lambda-2\sigma)(-1+2\gamma+2\lambda-2\sigma)  \\
\nonumber
&\times&  {}_2F_1\left(\dfrac{3}{2}-\gamma-\lambda+\sigma,\dfrac{3}{2}-\gamma +\lambda+\sigma;2\,(1+\sigma);\dfrac{1}{1+e^{-aL}}\right) \Bigg]\Bigg\}, 
\nonumber\\\\
\phi_{3R}(0) &=& - i\, b_{1} \,\left(\dfrac{1}{1+e^{-aL}}\right)^\sigma \left(1-\dfrac{1}{1+e^{-aL}}\right)^{-\gamma} \left(E+\dfrac{V_0}{1+e^{-aL}}\right)  \nonumber\\
& \times&   {}_2F_1\left(\dfrac{1}{2} - \gamma-\lambda+\sigma, \dfrac{1}{2} - \gamma+\lambda+\sigma; 1+2\sigma; \dfrac{1}{1+e^{-aL}}\right).
\end{eqnarray}

\bigskip
Subsequently, equating each component of the equations, we obtain the following relation,

\vspace{-0.5cm}
\begin{equation}
\label{41}
a_1 = b_1.
\end{equation}

This allows us to get the algebraic equation below, which we can use to determine the energy $E$ values,

\begin{eqnarray}
\label{eq42}
2\left(1+e^{-a L}\right) \sigma &+& \dfrac{1}{\left(1+2\sigma\right)}\Bigg[\dfrac{\left(\dfrac{1}{2} -\gamma-\lambda+\sigma\right)\left(\dfrac{1}{2}-\gamma+\lambda+\sigma\right) 
}{{}_2F_1\left(\dfrac{1}{2} - \gamma-\lambda+\sigma, \dfrac{1}{2} - \gamma+\lambda+\sigma; 1+2\sigma; \dfrac{1}{1+e^{-a L}}\right)} \nonumber\\ 
& \times& {}_2F_1\left(\dfrac{3}{2}-\gamma-\lambda+\sigma,\dfrac{3}{2}-\gamma +\lambda+\sigma;2\,(1+\sigma);\dfrac{1}{1+e^{-a L}}\right) \nonumber\\
\nonumber
& +& \ \dfrac{\left(\dfrac{1}{2} +\gamma-\lambda+\sigma\right)\left(\dfrac{1}{2}+\gamma+\lambda+\sigma\right)}{{}_2F_1\left(\dfrac{1}{2} + \gamma-\lambda+\sigma, \dfrac{1}{2} + \gamma+\lambda+\sigma; 1+2\sigma; \dfrac{1}{1+e^{-a L}}\right)}\\
&\times& {}_2F_1\left(\dfrac{3}{2}+\gamma-\lambda+\sigma,\dfrac{3}{2}+\gamma +\lambda+\sigma;2\,(1+\sigma);\dfrac{1}{1+e^{-a L}}\right)\Bigg] = 0.
\nonumber
\\
\end{eqnarray}

Using Eq. (\ref{eq42}), we can represent the energy  $E$ eigenvalues of the bound states for the DKP equation \cite{rojas:2024} with the Woods--Saxon potential well. This equation is the same as we get when solving for the Klein--Gordon equation for a Woods--Saxon potential well \cite{rojas2006klein}.

\subsection{Critical Conditions for Pair Creation}
A clear physical interpretation of the critical conditions for pair creation in the relativistic Klein?Gordon framework with a one--dimensional Woods?Saxon potential well is presented in \cite{rojas2006klein}. The authors solve the Klein--Gordon equation analytically and derive explicit expressions for the bound states. They show that as the potential depth increases, a bound antiparticle state emerges and eventually merges with the particle state at a critical potential $V_0 = V_{cr}$, where the norm of the wave function vanishes. This \textit{turning point} marks the onset of the supercritical regime, associated with vacuum instability and pair creation. The behavior is illustrated in Eq. (17) and Figures~2--5 of \cite{rojas2006klein}.

The normalization of the norm $N$ of the wave functions (\ref{bs_eq059}) and (\ref{bs_eq060}) is given by \cite{greiner:1985},

\begin{equation}
N= 2 \int_{-\infty}^{\infty}  [E-V(x)] \,\phi^*_{LR}(x) \phi_{LR}(x) \ \D x,
\label{eqnorm}
\end{equation}
where $\phi^*_{LR}(x)$ represents the conjugate of the wave function. The energy $E$ of the particle bound states $(E^{(+)})$ and the antiparticle bound states $(E^{(-)})$ correspond to $N>0$ and $N<0$, respectively. Furthermore, for $N=0$ both solutions meet and have the same energy $E$, which holds for all variations in the parameter $a$ \cite{rojas2006klein}. We can expand Eq. \eqref{eqnorm} in both regions of the Woods--Saxon potential well, in the following way,

\begin{equation}
N= 2 \int_{-\infty}^{0}  [E-V(x)]\, \phi^*_{L}(x) \phi_{L}(x) \ \D x + 2 \int_{0}^{\infty}  [E-V(x)] \,\phi^*_{R}(x)\phi_{R}(x) \ \D x,
\label{eqnorm_1}
\end{equation}
where $V(x)$ is given by  Eq. \eqref{eq002_well}.

\subsection{Bound states spectrum for Woods--Saxon potential well}

The point where the norm $N$ is equal to zero is called the turning point $V_{cr}$. The energy values $E^{(\pm)}$ for the particle and antiparticle bound states are calculated by numerically finding the roots of the eigenvalue equation \eqref{eq42}, increasing the value of $V_0$ in small steps, and using the Newton method implemented in the \texttt{FindRoot} command of Wolfram Mathematica\textsuperscript{\tiny\textregistered} \cite{Mathematica}. The root--finding procedure is initialized near each expected solution, which is located graphically. The critical value $V_{cr}$ is determined by incrementally adjusting $V_0$ until both energy solutions $E^{(\pm)}$ meet. A possible source of numerical error arises when searching for roots in the vicinity of the critical potential $V_{cr}$, where the solutions become highly sensitive to small changes in the depth of the Woods--Saxon potential well, so we use the Residual Error implemented in the \texttt{FindRoot} command which  evaluates the Eq. \eqref{eq42} at the point $V_{cr}$ to see how close it is to zero. In this sense, we show in Table \ref{table:turning_points_1} the value of $V_{cr}$, its corresponding Residual Error, the energy $E^{(\pm)
}$, and the norm $N$ for each case. The calculation of the norm $N$ is made using Eq. \eqref{eqnorm_1} for values near the critical potential $V_{cr}$, here we only show the values for the particle bound states.  The calculation of the norm of the wave function was performed using the energy $E$ values obtained by solving the eigenvalue problem of the system. These values were calculated for different values of the depth of the potential well $V_0$, with the aim of analyzing the behavior of the norm $N$ as a function of energy $E$ and determining the appearance of particle--antiparticle pairs.

The norm $N$ changes sign depending on the energy region, even for the same value of $V_0$. This change indicates a transition in the energy spectrum, associated with the disappearance of bound states and their connection to the continuum of negative energies. This phenomenon allows for the identification of the critical potential $V_{cr}$, defined as the value of $V_0$ at which the transition occurs.

\begin{figure}[htbp!]
\centering
\includegraphics[scale=0.45]{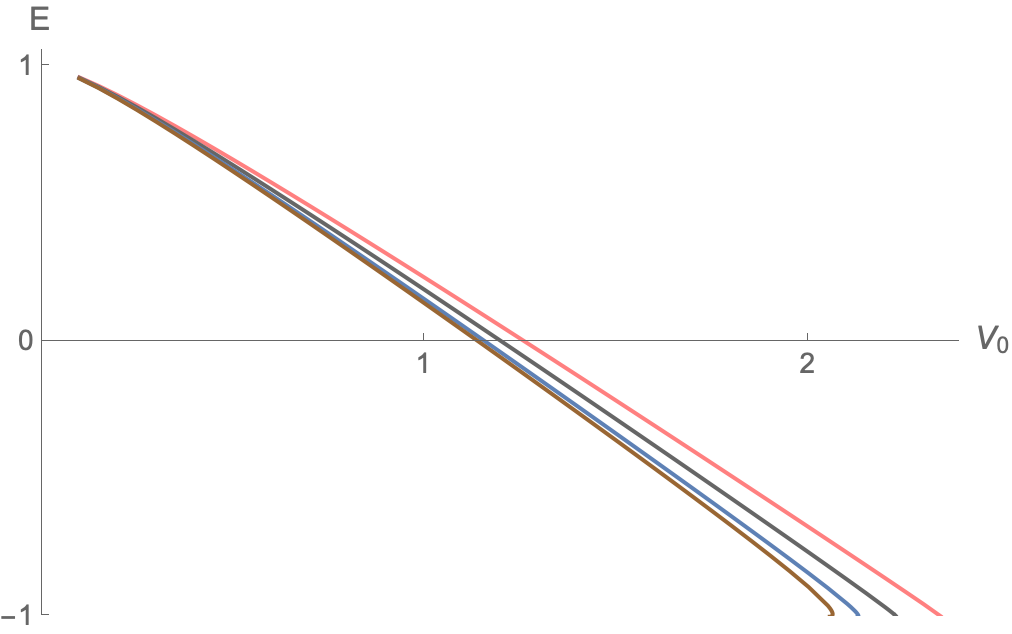}
\caption{Energy $E$ versus the depth of the Woods--Saxon potential well with $L = 2$,  and several values of $a$ in function of $V_0$. The solid pink line represents $a=2$, the solid gray line represents $a=3$, the solid blue line represents $a=6$, and the solid brown line represents $a=18$. In this figure we can only observed the turning point for the value of $a=18$.}
\label{FigL2}
\end{figure}

\begin{figure}[htbp!]
\centering
\includegraphics[scale=0.4]{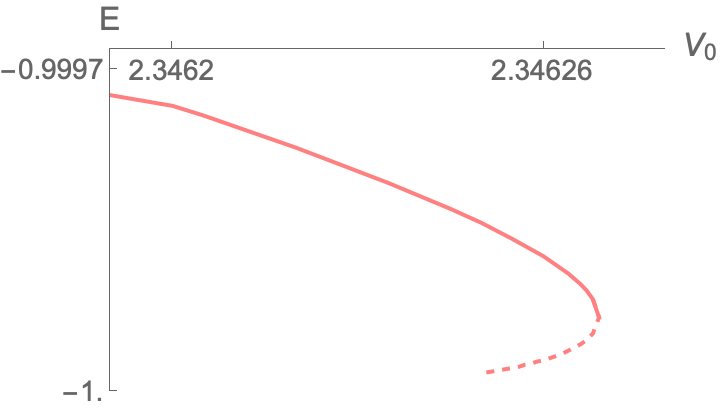}
\includegraphics[scale=0.4]{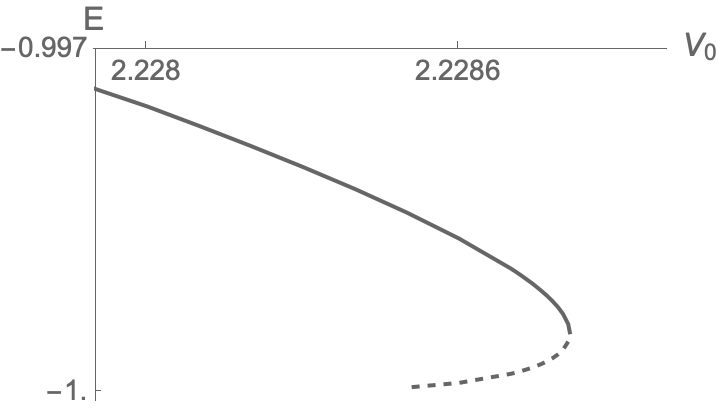}
\includegraphics[scale=0.4]{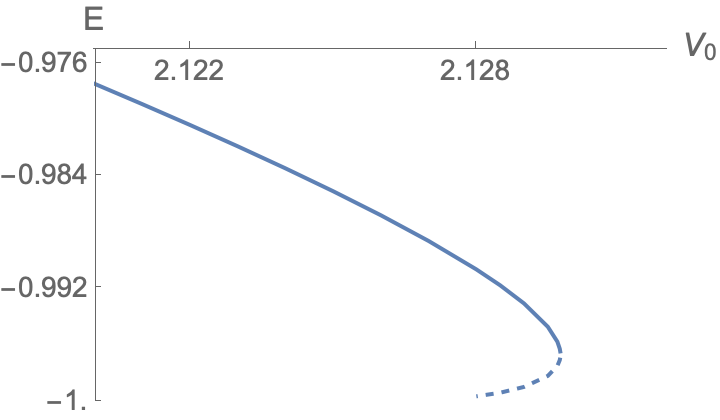}
\includegraphics[scale=0.4]{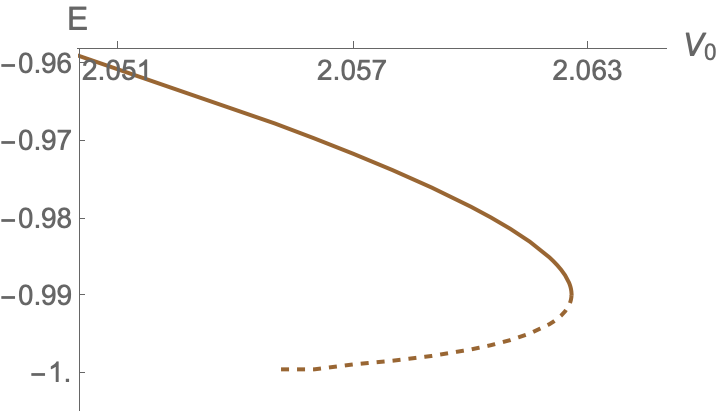}
\caption{Enlargement of the critical region where a turning point appears with zero norm $N$ for the particle and
antiparticle bound states. The solid line represents the particle bound states, and the dashed line represents the antiparticle bound states. In this figure we  observed the turning point for all considered values of $a$, this implies that we can more clearly observe the appearance of antiparticle bound states.}
\label{FigaL2_zoom_L}
\end{figure}

Figure \ref{FigL2} shows the behavior of the energy $E$ versus $V_0$ for several shapes of the Woods--Saxon potential well considering $L=2$, and $a=2$, $a=3$, $a=6$, and $a=18$.  Figure \ref{FigaL2_zoom_L} shows the region where the antiparticle arises. When $N=0$, there are two values of energy $E$ for one value of the depth of the potential well $V_0$.

\begin{table}[htbp!]
\centering
\caption{Norm $N$ of the critical potential $V_{cr}$ for the Woods--Saxon potential well with $L=2$, and several values of $a$.}
\vspace{0.5cm} 
\begin{tabular}{|c|c|c|c|c|} 
\toprule
\textbf{$a$} & \textbf{$V_{cr}$} & \textbf{ $E^{(\pm)}$}  & \textbf{ $N$} & \textbf{Residual Error for $V_{cr}$}\\ 
\midrule
2 & 2.34627 & -0.99993 & -1.30922 $\times 10^{-1}$ & 1.05365 $\times 10^{-8}$\\ 
\hline
3 & 2.22881 & -0.99949 & 1.11673  $\times 10^{-1}$ &2.33708 $\times 10^{-9}$\\ 
\hline
6 & 2.12976 & -0.99669 & 3.10581 $\times 10^{-2}$& 1.26438 $\times 10^{-9}$\\ 
\hline
18 &  2.06256 & -0.98971 & -2.21645 $\times 10^{-3}$ & $\mathcal{O}(10^{-8})$\\ 
\bottomrule
\end{tabular}
\label{table:turning_points_1}
\end{table}

In the following plots, we want to show the behavior of the bound states solutions for the Woods--Saxon potential well by considering $a=70$, and four different values of $L$. In Table \ref{table:turning_points_1.2} we  present the same calculation for Table \ref{table:turning_points_1}. In this case we also calculate the norm $N$ for  values near of the critical potential $V_{cr}$, where we   show values for particle and antiparticle bound states.

\begin{figure}[htbp!]
\centering
\includegraphics[scale=0.6]{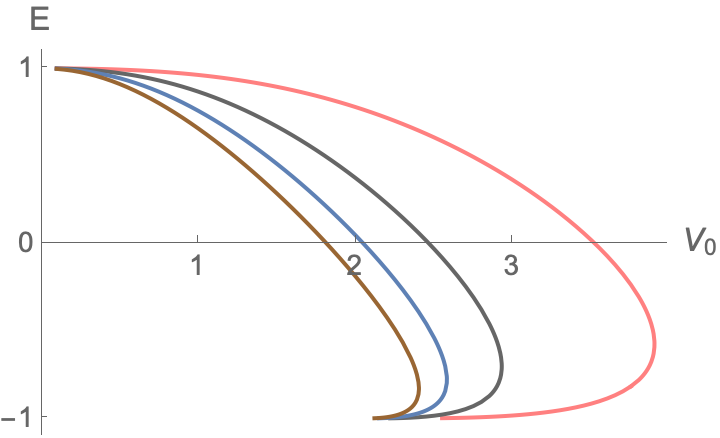}
\caption{Energy $E$ versus the depth of the Woods--Saxon potential well with $a = 70$,  and several values of $L$ in function of $V_0$. The solid pink line represents $L=0.1$, the solid gray line represents $L=0.2$, the solid blue line represents $L=0.3$, and the solid brown line represents $L=0.4$. This figure shows the turning point for all considered values of $a$, without requiring a zoom into the critical region where the antiparticle bound states emerge.}
\label{Figa70}
\end{figure}

\begin{figure}[htbp!]
\includegraphics[scale=0.45]{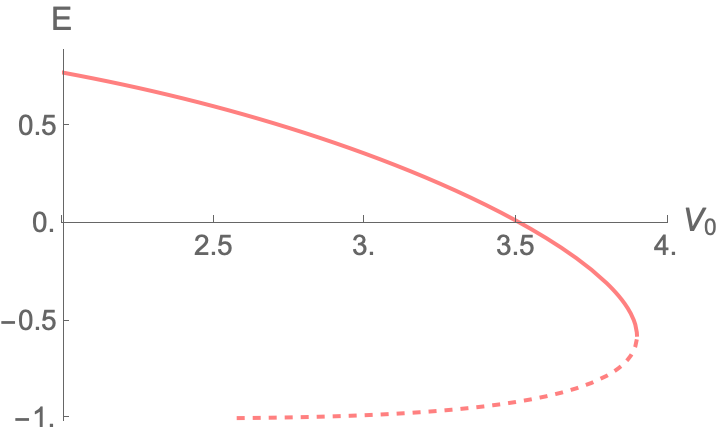}
\includegraphics[scale=0.45]{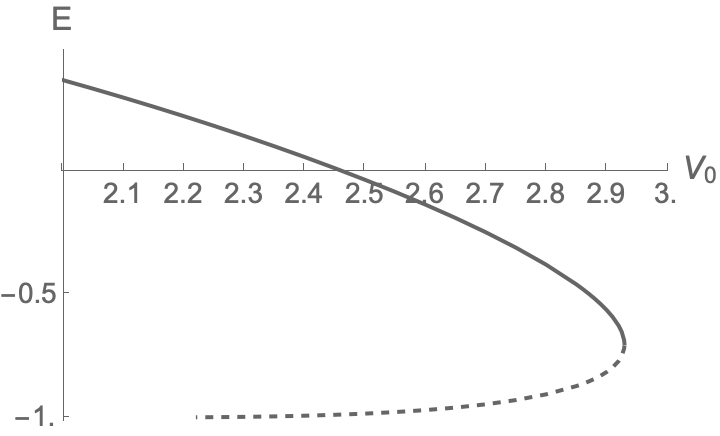}
\includegraphics[scale=0.45]{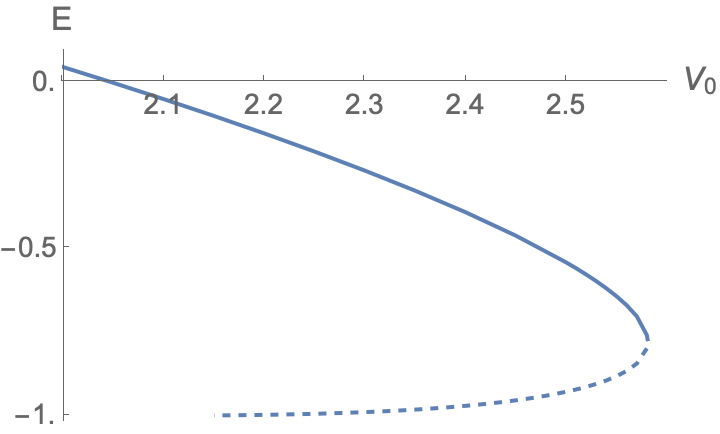}
\includegraphics[scale=0.45]{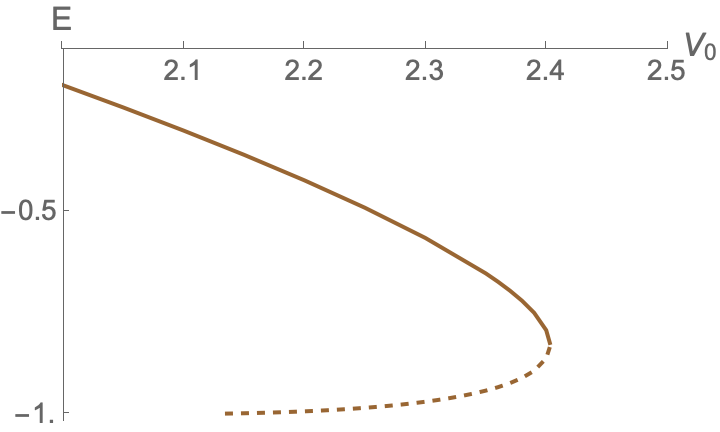}
\caption{Enlargement of the critical region where a turning point appears with zero norm $N$ for the particle--antiparticle bound states. The solid line represents the particle bound states, and the dashed line represents the antiparticle bound states. This figures show more clearly the turning point for all considered values of $a$, as a result, the visibility of antiparticle bound states is enhanced.}
\label{FigaL2_zoom_a}
\end{figure}

\begin{table}[htbp!]
\centering
\caption{Norm $N$ of the critical potential $V_{cr}$ for the Woods--Saxon potential well with $a=70$, and several values of $L$.}
\vspace{0.5cm} 
\begin{tabular}{|c|c|c|c|c|} 
\toprule
\textbf{$L$} & \textbf{$V_{cr}$} & \textbf{ $E^{(\pm)}$}  & \textbf{ $N$} & \textbf{Residual Error for $V_{cr}$}\\  
\midrule
0.1 & 3.90056 & -0.57536 & 1.31212 $\times 10^{-3}$ & $4.57925\times 10^{-13}$\\ 
\hline
0.2 & 2.92953 & -0.70249 & 3.91291$\times 10^{-3}$ &2.10959 $\times 10^{-11}$\\ 
\hline
0.3 & 2.58084 & -0.77846 & 5.24853 $\times 10^{-3}$&1.09252 $\times 10^{-9}$\\ 
\hline
0.4 &  2.40307 & -0.82924 & 4.49504 $\times 10^{-3}$ & 3.72292 $\times 10^{-4}$ \\ 
\bottomrule
\end{tabular}
\label{table:turning_points_1.2}
\end{table}

\newpage
\subsection{Pair creation in the square potential well vs. the Woods--Saxon potential well}

\begin{figure}[th!]
\centering
\includegraphics[scale=0.55]{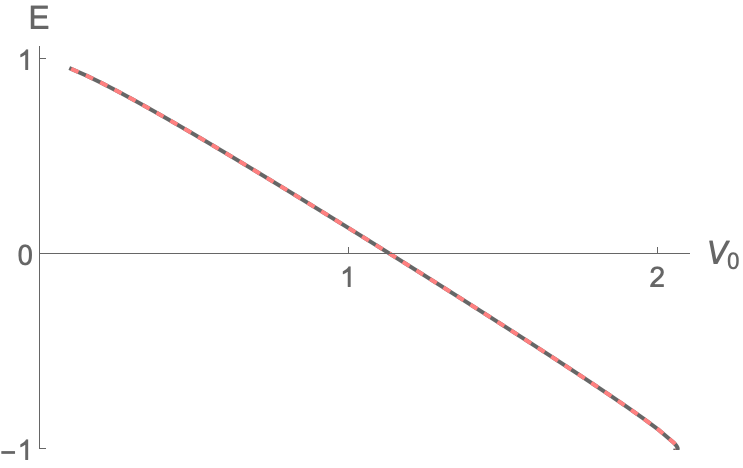}
\caption{The bound states spectrum for a square potential well  with $L=2$, and the Woods--Saxon potential well with $L=2$, and $a=18$. The gray solid line represents the square potential well, and the dashed pink line the Woods--Saxon potential well. This figures show clearly the appear of the critical region for both potential wells.}
\label{FigswandwsL2}
\end{figure}

Figure \ref{FigswandwsL2} illustrates the bound states spectra for both potential wells. The plot on a gray solid line corresponds to the square potential well, while the plot with a dashed pink line represents the Woods--Saxon potential well. Both potential  wells have a width of $L = 2$, and in the case of the Woods--Saxon potential well, a steepness parameter of  $a = 18$  is chosen because it closely approximates the shape of a square well. 

\begin{figure}[th!]
\centering
\includegraphics[scale=0.55]{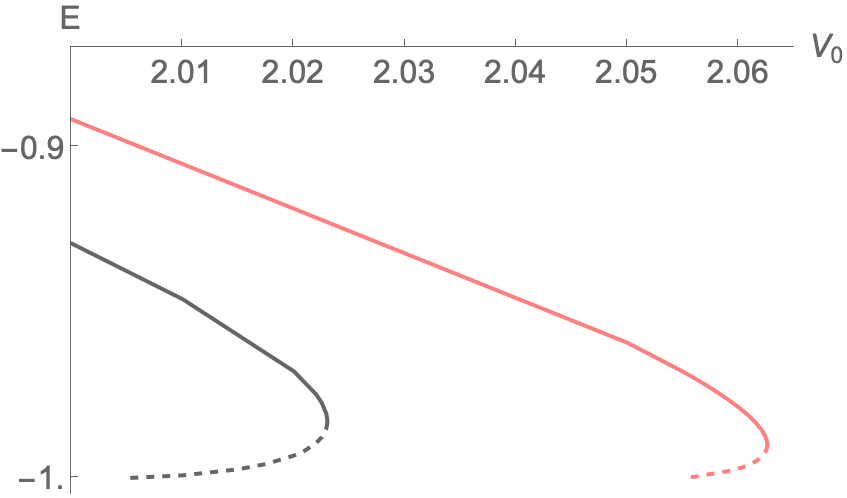}
\caption{Enlargement of the critical region where a turning point appears with zero norm $N$ for the particle and antiparticle bound states. The square potential well is plotted considering $L=2$, and the Woods--Saxon potential well with $L=2$ and $a=18$. The solid--dashed gray line represents the square potential well, and the solid--dashed pink line represents the Woods--Saxon potential well. In this figure, antiparticle bound states become more clearly observable for both potential wells.}
\label{FigswandwsL2_ZOOM}
\end{figure}

At first glance, both potential wells look similar, as the Woods--Saxon potential well with a high steepness parameter $a = 18$ closely resembles the sharp boundaries of the square potential well.   Table \ref{table:turning_points_2} shows the value of the critical potential $V_{cr}$, the energy $E^{(\pm)}$, the norm $N$, and the Residual Error for $V_{cr}$ for both potential wells, this calculations were only make for  particle bound states. We can see that in both potential wells we have pair creation. In Figure \ref{FigswandwsL2_ZOOM} we made a zoom of the turning point region, and we can observe that the value of $V_{cr}$  differs a little because we need to use a larger value of $a$ in order to get the Woods--Saxon potential well closer to the square potential well. When $V_0$ reaches $V_{cr}$, the bound states merge with this continuum, facilitating the creation of particle--antiparticle pairs. Unlike the square potential well, which exhibits abrupt transitions, the Woods--Saxon potential well, being continuous and smooth, shows a more gradual transition between bound states and the continuum. This facilitates the precise identification of $V_{cr}$ and the analysis of the energy spectrum.

\begin{table}[htbp!]
\centering
\caption{Norm $N$ of the critical potential $V_{cr}$ for the square potential well with $L=2$ and the Woods--Saxon potential well with $L=2$, and $a=18$.}
\scalebox{0.9}{
\begin{tabular}{|c|c|c|c|c|} 
\toprule
\textbf{Potential well} & \textbf{$V_{cr}$} & \textbf{ $E^{(\pm)}$}  & \textbf{ $N$} & \textbf{Residual Error for $V_{cr}$}\\ 
\midrule
Square & 2.02299 & \begin{tabular}{c} -0.98269 \\-0.98257\end{tabular}& \begin{tabular}{c}-8.82573 $\times 10^{-3}$ \\7.83672 $\times 10^{-3}$\end{tabular} & \begin{tabular}{c} 5.55112 $\times 10^{-17}$ \\6.10623 $\times 10^{-16}$\end{tabular}\\ 
\midrule
Woods--Saxon & 2.06256 & \begin{tabular}{c} -0.98968 \\-0.98978\end{tabular}& \begin{tabular}{c} 2.19475 $\times 10^{-2}$ \\-5.89707 $\times 10^{-2}$\end{tabular}& \begin{tabular}{c} $\mathcal{O}(10^{-8})$\\$\mathcal{O}(10^{-8})$\end{tabular}\\
\bottomrule
\end{tabular}
}
\label{table:turning_points_2}
\end{table}

\subsection{Pair creation in the cusp potential well vs. the Woods--Saxon potential well}

\begin{figure}[ht!]
\centering
\includegraphics[scale=0.5]{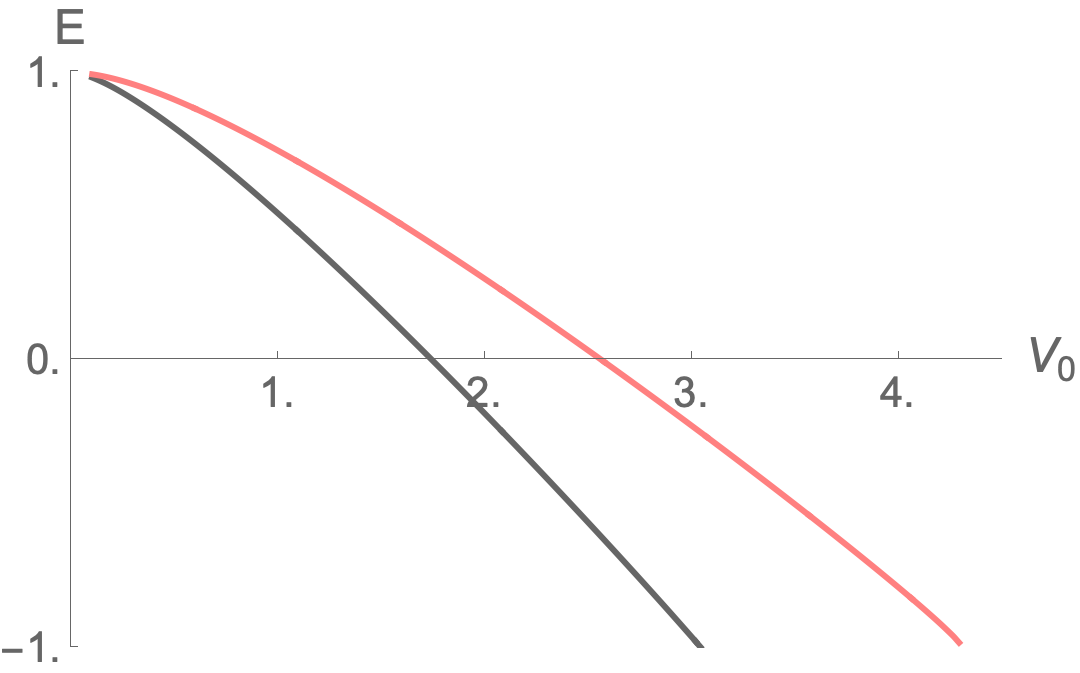}
\caption{The bound states spectrum for a cusp potential well  with $a=1.2$, and the Woods--Saxon potential well with $a=2$, and $L=0.4$. The gray solid line represents the cusp potential well, and the solid pink line the Woods--Saxon potential well. In these figures, the critical region is not clearly visible for either potential well.}
\label{figscuspandws}
\end{figure}

\begin{figure}[ht!]
\includegraphics[scale=0.3]{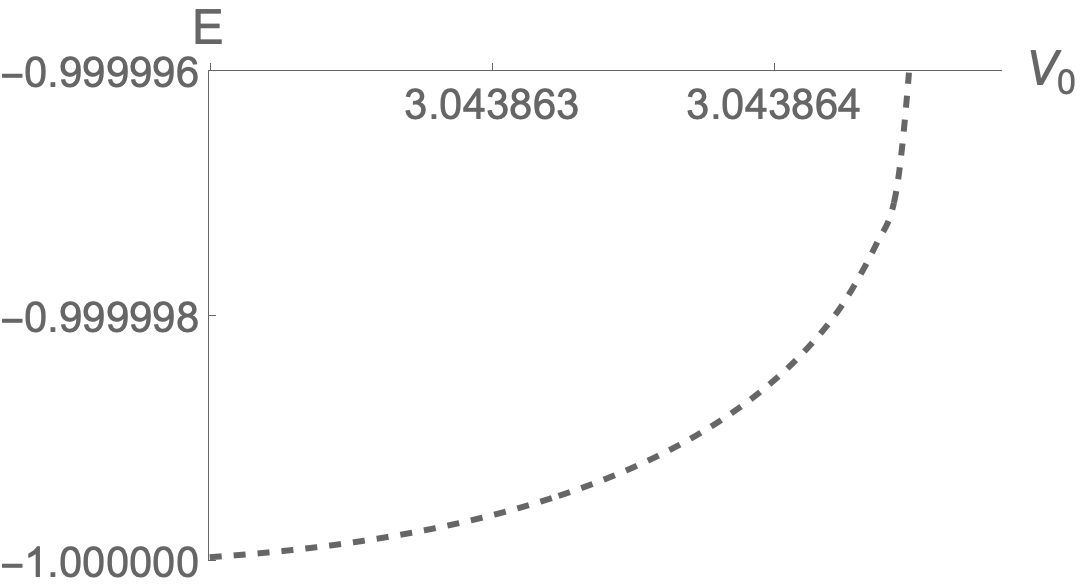}
\includegraphics[scale=0.3]{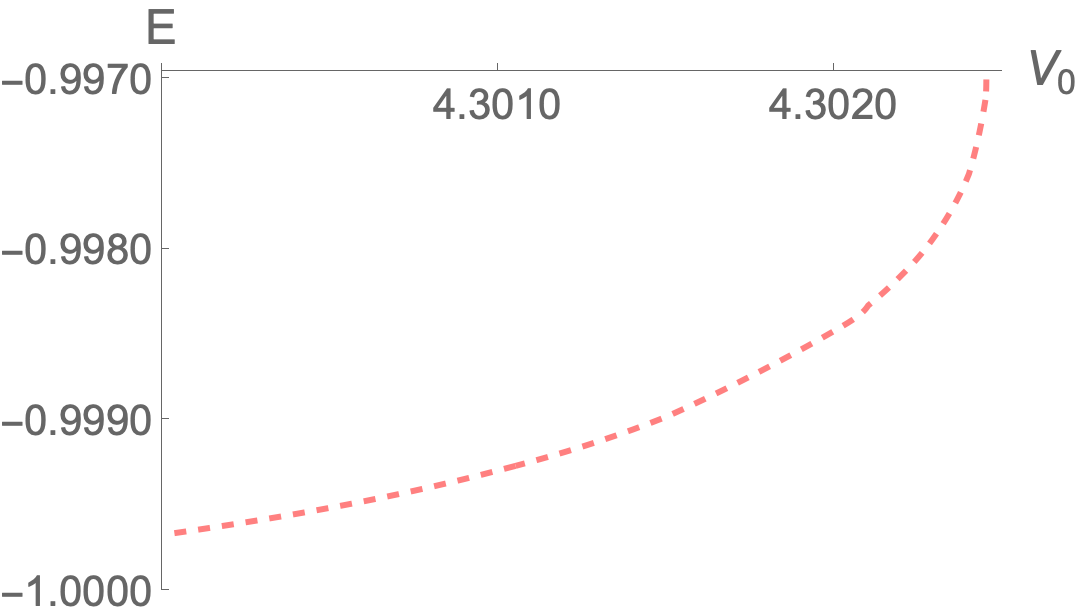}
\caption{Enlargement of the critical region where a turning point appears with zero norm $N$ for the particle and antiparticle bound states. (a) The cusp potential well is plotted considering $a=1.2$, and (b) the Woods--Saxon potential well considering $a=2$, and $L=0.4$. The dashed gray line represents the cusp potential well, and the dashed pink line represents the Woods--Saxon potential well. This figure illustrates that the antiparticle bound states are more distinctly visible in both potential wells.}
\label{figcuspWS_ZOOM}
\end{figure}

Figure \ref{figscuspandws} illustrates the bound states spectra for the cusp potential well with $a=1.2$, and the Woods--Saxon potential well with $a=2$, and $L=0.4$.
With these parameters, both potential wells appear very similar, as the Woods--Saxon potential well with $a=2$, and $L=0.4$ closely resembles the peak structure of the cusp potential well.  Table \ref{table:turning_points_3} shows the value of the critical potential $V_{cr}$, the energy $E^{(\pm)}$, the norm $N$, and the Residual Error for $V_{cr}$ for both potentials, in this case we have particle bound states and the value of the norm $N$ for the cusp potential well presents problems with the integration method, then we only report the sign of the norm.  In Figure \ref{figcuspWS_ZOOM} we made a zoom of the turning point region, and we can observe that the value of $V_{cr}$ differs because we need to use different sets of parameters to get the Woods--Saxon potential well closer to the cusp potential well.

\begin{table}[htbp]
\centering
\caption{Norm $N$ of the critical potential $V_{cr}$ for the cusp potential well with $a=1.2$,  and the Woods--Saxon potential well with $a=2$, and $L=0.4$}
\scalebox{0.98}{
\begin{tabular}{|c|c|c|c|c|} 
\toprule
\textbf{Potential well} & \textbf{$V_{cr}$} & \textbf{ $E^{(\pm)}$}  & \textbf{ $N$} & \textbf{Residual Error for $V_{cr}$}\\  
\midrule
Cusp  & 3.04386 & -0.99999 &  $>0$ &6.70502$\times 10^{-9}$\\ 
\hline
Woods--Saxon  &4.30245  & -0.99690 &3.70029  $\times 10^{-1}$& 7.91763 $\times 10^{-9}$\\ 
\bottomrule
\end{tabular}
}
\label{table:turning_points_3}
\end{table}

\newpage
\section{Conclusions}	
\label{sec_conclusion}

This research analyzes the transmission resonances and the bound states solutions for a spatially one--dimensional Woods--Saxon potential barrier and well like solution to the relativistic Duffin--Kemmer--Petiau (DKP) equation. This algebraic process was performed for  $(1+1)$ dimensions, considering the equivalence between spin--one and spin--zero cases due to the $(3 \times 3)$ $\beta$ matrices, which define a three--component DKP spinor. 

Moreover, the scattering process allow us to analyze  the transmission $T$ and reflection $R$ coefficients, which were  derived from the solutions of the system of differential equations obtained from the spinors in DKP theory, employing the definition of $j^\mu$ currents. These solutions were expressed in terms of the Gaussian hypergeometric functions, the Gaussian regularized hypergeometric functions, and the Gamma $\Gamma(x)$ functions. Once calculated the transmission $T$ and reflection $R$ coefficients, we  study the transmission resonances for several values of the Woods--Saxon potential barrier, which were obtained in all considered cases, additionally we show that when $T=1$ then $R=0$, according to the unitary condition $T+R=1$. These results have been compared with those obtained for the square potential barrier and the cusp potential barrier, using two specific sets of parameters for the Woods--Saxon potential barrier to approach the corresponding limiting cases.
 
The bound states solutions were obtained by solving the DKP equation separately in the positive and negative regions of the potential well. The resulting solutions were expressed in terms of the Gaussian hypergeometric functions and their asymptotic behavior was analyzed to ensure regularity. Additionally, the other two components of the DKP spinor were determined for both regions of the potential well.

The norm $N$ was explicitly calculated in several cases to verify the conditions for bound states, distinguishing between particle bound states $(N>0)$ and antiparticle bound states $(N<0)$. First, we considered particle bound states solutions for the Woods--Saxon potential well by fixing $L=2$ and varying the parameter $a$ over four different values. Second, we examined particle bound states by fixing $a=70$ and varying $L$ over four different values.

Furthermore, the Woods--Saxon potential well was compared with the square potential well and the cusp potential well to examine the conditions for pair creation in the limiting cases of the Woods--Saxon potential well. The study confirmed that pair creation occurs in all of these potential wells. For these cases, the norm $N$ was calculated for both particle and antiparticle bound state solutions, first in the comparison between the square and Woods--Saxon potential wells, and then in the comparison with the cusp potential well.

In summary, this work provides a detailed understanding of transmission resonances in the Duffin--Kemmer--Petiau (DKP) equation with a Woods--Saxon potential barrier, as well as the particle and antiparticle energy spectrum. It also highlights the role of the critical potential $V_{cr}$ in the creation of particle--antiparticle pairs within the Woods--Saxon potential well in the framework of the DKP equation.


\section{Acknowledgments}	
\label{sec_acknowledgements}

\noindent
L. M. P. acknowledges financial support from ANID through Convocatoria
Nacional Subvenci\'on a Instalaci\'on en la Academia Convocatoria A\~no
2021, Grant No. SA77210040.\\

\noindent
D. L. acknowledges partial financial support from Centers of Excellence
with BASAL/ANID financing, Grant No. AFB220001, CEDENNA.

\bibliographystyle{unsrt}


\end{document}